\documentclass[10pt,journal,compsoc]{IEEEtran}
\ifCLASSOPTIONcompsoc
  \usepackage[nocompress]{cite}
\else
  \usepackage{cite}
\fi
\ifCLASSINFOpdf
\else
\fi
\usepackage{booktabs} 
\usepackage{subfig}
\usepackage{graphicx}
\usepackage{algorithm}
\usepackage{algpseudocode}%
\usepackage{xcolor}
\usepackage{amsmath}
\usepackage{comment}
\newcommand{\etal}{\textit{et al}. }
\usepackage{mathtools}
\usepackage{multirow}
\usepackage{graphics}
\usepackage{enumitem}
\usepackage{CJKutf8}
\usepackage{array}
\usepackage{diagbox}
\usepackage{hyperref}
\usepackage[utf8]{inputenc}
\usepackage{bm}

\usepackage{ragged2e}
\usepackage{amsmath,amssymb}

\usepackage{xcolor,cite,etoolbox}
\definecolor{mypurple}{rgb}{0.4392, 0.1882, 0.6275}
\definecolor{mygreen}{rgb}{0, 0.6902, 0.3137}
\makeatletter 
\pretocmd\@bibitem{\csname keycolor#1\endcsname}{}{\fail}
\newcommand\citecolor[1]{\@namedef{keycolor#1}{}}
\renewcommand{\algorithmicrequire}{\textbf{Input:}}

\begin{document}
%
\title{Semi-Supervised Learning for Multi-Label Cardiovascular Diseases Prediction:\\ A Multi-Dataset Study}
%
%
%
%
\author{Rushuang Zhou, Lei Lu, Zijun Liu, Ting Xiang, Zhen Liang, David A. Clifton, \\Yining Dong and Yuan-Ting Zhang,~\IEEEmembership{Fellow,~IEEE}
\IEEEcompsocitemizethanks{\IEEEcompsocthanksitem Rushuang Zhou, Zijun Liu, Ting Xiang are with the Department
of Biomedical Engineering, City university of Hong Kong, Hong Kong,
China and also with Hong Kong Center for Cerebro-Cardiovascular Health Engineering (COCHE), Hong Kong, China.
\IEEEcompsocthanksitem Lei Lu and David A. Clifton are with the Department of Engineering Science, University of Oxford, UK.
\IEEEcompsocthanksitem David A. Clifton is also with Oxford-Suzhou Institute of Advanced Research (OSCAR), Suzhou, China.
\IEEEcompsocthanksitem Zhen Liang is with the School of Biomedical Engineering, Medical School, Shenzhen University, Shenzhen, China.
\IEEEcompsocthanksitem Yining Dong is with the School
of Data Science, City University of Hong Kong, Hong Kong,
China and also with Hong Kong Center for Cerebro-Cardiovascular Health Engineering (COCHE),  Hong Kong, China.
\IEEEcompsocthanksitem Yuan-Ting Zhang is with Micro sensing and imaging technologies limited, Hong Kong, China and also with Wearable intelligent sensing technologies limited, Hong Kong, China.
\IEEEcompsocthanksitem Corresponding authors: Yining Dong and Yuan-Ting Zhang, e-mail: yinidong@cityu.edu.hk, ytzhanghicas@gmail.com. }

\thanks{Manuscript received April 19, 2005; revised August 26, 2015.}}

%
%

\markboth{Journal of \LaTeX\ Class Files,~Vol.~14, No.~8, August~2015}%
{Shell \MakeLowercase{\textit{et al.}}: Bare Demo of IEEEtran.cls for Computer Society Journals}
%



\IEEEtitleabstractindextext{%
\begin{abstract}
Electrocardiography (ECG) is a non-invasive tool for predicting cardiovascular diseases (CVDs). Current ECG-based diagnosis systems show promising performance owing to the rapid development of deep learning techniques. However, the label scarcity problem, the co-occurrence of multiple CVDs and the poor performance on unseen datasets greatly hinder the widespread application of deep learning-based models. Addressing them in a unified framework remains a significant challenge. To this end, we propose a multi-label semi-supervised model (ECGMatch) to recognize multiple CVDs simultaneously with limited supervision.  In the ECGMatch, an ECGAugment module is developed for weak and strong ECG data augmentation, which generates diverse samples for model training. Subsequently, a hyperparameter-efficient framework with neighbor agreement modeling and knowledge distillation is designed for pseudo-label generation and refinement, which mitigates the label scarcity problem. Finally, a label correlation alignment module is proposed to capture the co-occurrence information of different CVDs within labeled samples and propagate this information to unlabeled samples. Extensive experiments on four datasets and three protocols demonstrate the effectiveness and stability of the proposed model, especially on unseen datasets. As such, this model can pave the way for diagnostic systems that achieve robust performance on multi-label CVDs prediction with limited supervision.
\end{abstract}

}

\maketitle

\IEEEdisplaynontitleabstractindextext

%
\IEEEpeerreviewmaketitle

\IEEEraisesectionheading{\section{Introduction}\label{sec:introduction}}

%
%
%
%
\IEEEPARstart{C}{ardiovascular} diseases (CVDs) have become the world's leading cause of morbidity and mortality in recent years\cite{kelly2010promoting}. As a non-invasive test, the 12-lead electrocardiography (ECG) is widely used for diagnosing CVDs. With the rapid development of deep learning and artificial intelligence, AI-aided automatic diagnosis systems have attracted considerable interest in clinical practice. Most of these systems are designed for a well-defined setting where the annotated samples are sufficient and identically distributed, with each sample only belonging to one CVDs class. Unfortunately, the complex real-world setting differs from this ideal setting, where annotated ECG segments are tough to collect, and multiple CVDs can be identified from each segment. Furthermore, the real-world training and test data may not be sampled from the same distribution, which greatly hurts the model performance. The difference between the real-world setting and the ideal setting restricts the clinical applications of current systems. In a nutshell, there are three challenges in the clinical applications of automatic diagnosis systems: 1) Label scarcity problem. 2) Poor performance on unseen datasets. 3) Co-occurrence of multiple CVDs. 

In recent years, semi-supervised learning (SSL) has shown great potential in addressing the label scarcity problem in clinical applications. \textcolor{black}{The main idea of the SSL models is to utilize unlabeled samples for model training, which are easier to collect compared with labeled samples\cite{semi2006,lee2021abc}.} By leveraging the abundant information within the unlabeled samples, SSL models often outperform fully-supervised models when the number of labeled samples is limited\cite{berthelot2019mixmatch,sohn2020fixmatch,zhang2021flexmatch}. Consequently, numerous studies were proposed to extend the success of SSL to ECG-based CVDs prediction. For example, Oliveira \etal applied existing SSL models for ECG signal classification. Experiments on the MIT-BIH database\cite{moody2001impact} demonstrated the superiority of the SSL models compared with fully-supervised models\cite{oliveira2022generalizable}. To improve the model performance on unseen datasets, Feng \etal proposed a transfer learning framework to transfer the model trained on a label-sufficient
dataset to a label-scarce target dataset. Comprehensive results on four benchmarks demonstrated the robustness of the proposed framework. At the same time, multi-label learning sheds new light on how to detect multiple CVDs from one ECG recording simultaneously. In contrast to single-label learning, multi-label learning generates multiple predictions for a given sample, with each prediction indicating whether the sample belongs to a specific category\cite{liu2021emerging}.  Multi-label learning models have the capability to detect multiple diseases from ECG signals, while single-label models are limited to recognizing only one disease at a time. As a result, numerous models have been proposed to leverage multi-label learning for ECG-based CVDs prediction. For example, Strodthoff \etal evaluated the performance of existing models on the PTB-XL database\cite{wagner2020ptb,strodthoff2020deep} and demonstrated the feasibility of using multi-label learning models for CVDs prediction. Subsequently, Ge \etal and Ran \etal proposed utilizing the relationship between different cardiac diseases to enhance the model performance\cite{ge2021multi,ran2023label}.  More details about the existing models for ECG-based CVDs prediction are presented in Section \ref{sec:relatedWorksemi}. 

To the best of our knowledge, no prior study has proposed and validated a unified framework to tackle the aforementioned three challenges simultaneously. Specifically, most previous studies only alleviated one of the aforementioned problems without comprehensively considering the other two challenges in CVDs prediction. For example, many SSL-based models ignored the co-occurrence of multiple CVDs and their performance on unseen datasets was not satisfying\cite{zhang2021flexmatch,chen2023softmatch,zhai2020semi,zhang2022semi,oliveira2022generalizable}. Previous multi-label learning models could detect multiple CVDs from ECG signals, but their effectiveness relied heavily on sufficient labeled data and handcrafted prior knowledge\cite{strodthoff2020deep, ge2021multi,ran2023label}. These significant deficiencies imply that these models are still far from being applicable in real-world scenarios. Therefore, in this study, we propose a multi-label semi-supervised framework (ECGMatch) that can use only 1\% of the annotated samples to achieve good results in cross-dataset multi-label CVDs prediction. Here, we introduce how the proposed framework addresses the aforementioned problem simultaneously.

First, a novel ECGAugment module is developed to alleviate the label scarcity problem by generating diverse samples. It exploits the intrinsic characteristics of ECG signals and dramatically outperforms traditional methods\cite{kiyasseh2021clocs,oliveira2022generalizable}. Moreover, we design a pseudo-label generation module that utilizes the interaction between the student and teacher networks to generate pseudo-labels for the unlabeled samples. Specifically, we formulate the generation task as a knowledge distillation process. During training, the teacher stores the learned knowledge in two memory banks, and the student visits the banks to assign pseudo-labels for the unlabeled samples using a K-Nearest voting strategy. To mitigate the negative impact of inaccurate pseudo-labels, we propose a neighbor agreement modeling method and develop a hyper-parameter efficient module for refining these labels. During the K-Nearest voting process, the degree of agreement among neighbors can be utilized to estimate the pseudo-label confidence, which is an important indicator for discovering truth-worthy pseudo-labels. In multi-label learning, the advantage of the proposed hyperparameter-efficient refinement module is more significant as it only relies on the number of neighbors $K$ rather than numerous thresholds and complex control strategies\cite{sohn2020fixmatch,zhang2021flexmatch,rizve2021defense,chen2023softmatch}.

To capture the co-occurrence of different CVDs, we introduce a label correlation alignment module. It quantitatively estimates the co-occurrence information using limited labeled data and transfers this knowledge to unlabeled data. In practice, we compute a correlation matrix to represent the co-occurrence information, and align the matrices computed on labeled and unlabeled data to complete a knowledge transfer process. Finally, we conduct extensive experiments on four public datasets across three protocols. The results comprehensively validate the superiority of the ECGMatch, especially on unseen datasets. In summary, the main contributions and novelties are listed below. 
\begin{itemize}
\item We proposed a robust pipeline for ECG signal augmentation, which shows remarkable improvements compared with previous methods. 
\item An efficient method for pseudo-label refinement is developed for multi-label learning with limited supervision. It has fewer parameters than threshold-based methods but shows better performance. 
\item A novel approach is proposed to align the label correlation computed on labeled and unlabeled data, which provides a reliable solution to capture the co-occurrence of multiple CVDs.
\item A unified semi-supervised framework for multi-label CVDs prediction is proposed, which is the first one to address three critical challenges in this area. 
\end{itemize}
\section{Related Work}
\label{sec:relatedWorksemi}
\subsection{ECG-Based CVDs Prediction Using Deep Learning}
Over the past decade, the potential and feasibility of utilizing ECG signals to diagnose a wide spectrum of CVDs have been demonstrated by numerous previous studies\cite{kiranyaz2015real,clifford2017af,hannun2019cardiologist,ribeiro2020automatic,strodthoff2020deep,kiyasseh2021clocs,kiyasseh2021clinical,jin2021novel,tesfai2022lightweight,huang2022snippet,huang2022novel}. With the rapid development of deep learning techniques, many studies used end-to-end deep learning models to achieve accurate predictions of the CVDs. For example, Kiranyaz \etal designed a real-time one-dimensional convolutional neural network (CNN) that achieved superior performance in ECG-based CVDs prediction compared with traditional models\cite{kiranyaz2015real}.  Hannun \etal conducted a comprehensive evaluation of a deep neural network (DNN) for ECG signal classification. The extensive results showed that the DNN model achieved a similar diagnosis performance to cardiologists, thus demonstrating its enormous potential in clinical applications\cite{hannun2019cardiologist}. Subsequently, several methods were proposed to enhance the accuracy of the DNN model. For example, Ribeiro \etal proposed a unidimensional residual neural network architecture that outperformed cardiology resident medical doctors in recognizing six kinds of CVDs\cite{ribeiro2020automatic}. Huang \etal introduced a novel deep reinforcement learning framework called snippet policy network V2 (SPN-V2) for the early prediction of CVDs based on ECG signals. Using a novel keen-guided neuroevolution algorithm, the SPN-V2 network achieved a stable balance between recognition accuracy and earliness\cite{huang2022snippet}. However, despite the significant advancements in ECG-based CVDs prediction using deep learning methods in recent years, such methods may experience a performance drop when the number of labeled samples is limited\cite{zhang2022semi}.

\subsection{Semi-Supervised Learning for ECG-Based CVDs Prediction.}
Semi-supervised learning has achieved great success in reducing the requirements on laborious annotations for model training\cite{berthelot2019mixmatch,sohn2020fixmatch,UDA2020,zhang2021flexmatch,chen2023softmatch}. As a result, an increasing number of studies have proposed using SSL to develop robust models for ECG-based CVDs prediction with limited supervision\cite{zhai2020semi,zhang2022semi,oliveira2022generalizable}. For instance, Zhai \etal proposed a semi-supervised model to transfer knowledge learned from large datasets to small datasets. Extensive experiments demonstrated that the performance of the proposed model was comparable to other methods which required numerous annotated samples\cite{zhai2020semi}. Oliveira \etal applied different SSL models for ECG-based CVDs prediction, such as MixMatch\cite{berthelot2019mixmatch} and FixMatch\cite{sohn2020fixmatch}. When only 15\% of the ECG data was labeled, the SSL models achieved comparable prediction performance obtained by fully supervised models\cite{oliveira2022generalizable}. Motivated by the mean teacher algorithm\cite{tarvainen2017mean}, Zhang \etal proposed the mixed mean teacher model for automatic atrial fibrillation detection using ECG, which significantly reduced the workload of data annotation by 98\% while achieving comparable performance as fully supervised models\cite{zhang2022semi}. To address the distribution shifts across different datasets, Feng \etal proposed a SSL framework based on two complementary modules: semantic-aware feature alignment (SAFA) and prototype-based label propagation (PBLP)\cite{feng2023semantic}. Comprehensive experiments verified that the proposed model achieved promising performance on target datasets using limited labeled target samples.

However, previous SSL studies for ECG-based CVDs prediction have two main limitations. \textbf{1) Previous SSL studies developed single-label classification models for CVDs prediction, which were greatly limited in clinical applications.} Specifically, they simply formulated the CVDs prediction task as a single-label problem, where each ECG signal can only belong to one category. However, multiple CVDs, such as atrial fibrillation and right bundle branch block, usually co-occur in one ECG segment\cite{ran2023label}. This phenomenon suggests that the CVDs prediction task should be formulated as a multi-label problem, where each ECG signal belongs to multiple categories.   
\textbf{2) Previous studies did not consider CVDs prediction on unseen datasets.} The training and test data in previous studies were from the same dataset, which is often unrealistic in real-world applications. While some studies applied transfer learning to transfer knowledge from the training datasets to unseen datasets\cite{strodthoff2020deep,feng2023semantic}, their methods still needed labeled samples from the test dataset, which led to information leaking.

\subsection{Multi-Label Model for ECG-Based CVDs Prediction. } Many studies have investigated the feasibility of using multi-label learning to simultaneously detect multiple arrhythmia types from ECG signals. Strodthoff \etal proposed a pilot study that evaluated the performance of different models for ECG-based multi-label CVDs classification and found that ResNet and Inception-based CNN architectures achieved the best performance\cite{strodthoff2020deep}. Ge \etal utilized Bayesian conditional probability to capture the association between ECG abnormalities and used it to guide the feature fusion of ECG-based models. Promising experimental results showed that the multi-label correlation guided feature fusion network outperformed other competitors\cite{ge2021multi}.  Ran \etal proposed a label correlation embedding guided network (LCEGNet) to capture the relationship between different ECG abnormalities and improve the model performance by learning arrhythmia-specific features\cite{ran2023label}. However, annotating multi-label ECG data is prohibitively expensive, leading to a critical bottleneck in real-world applications. This problem highlights the urgent need for semi-supervised learning in ECG-based multi-label CVDs prediction, which will be investigated in our study.
\begin{figure*}[h]
\begin{center}
\includegraphics[width=1\textwidth]{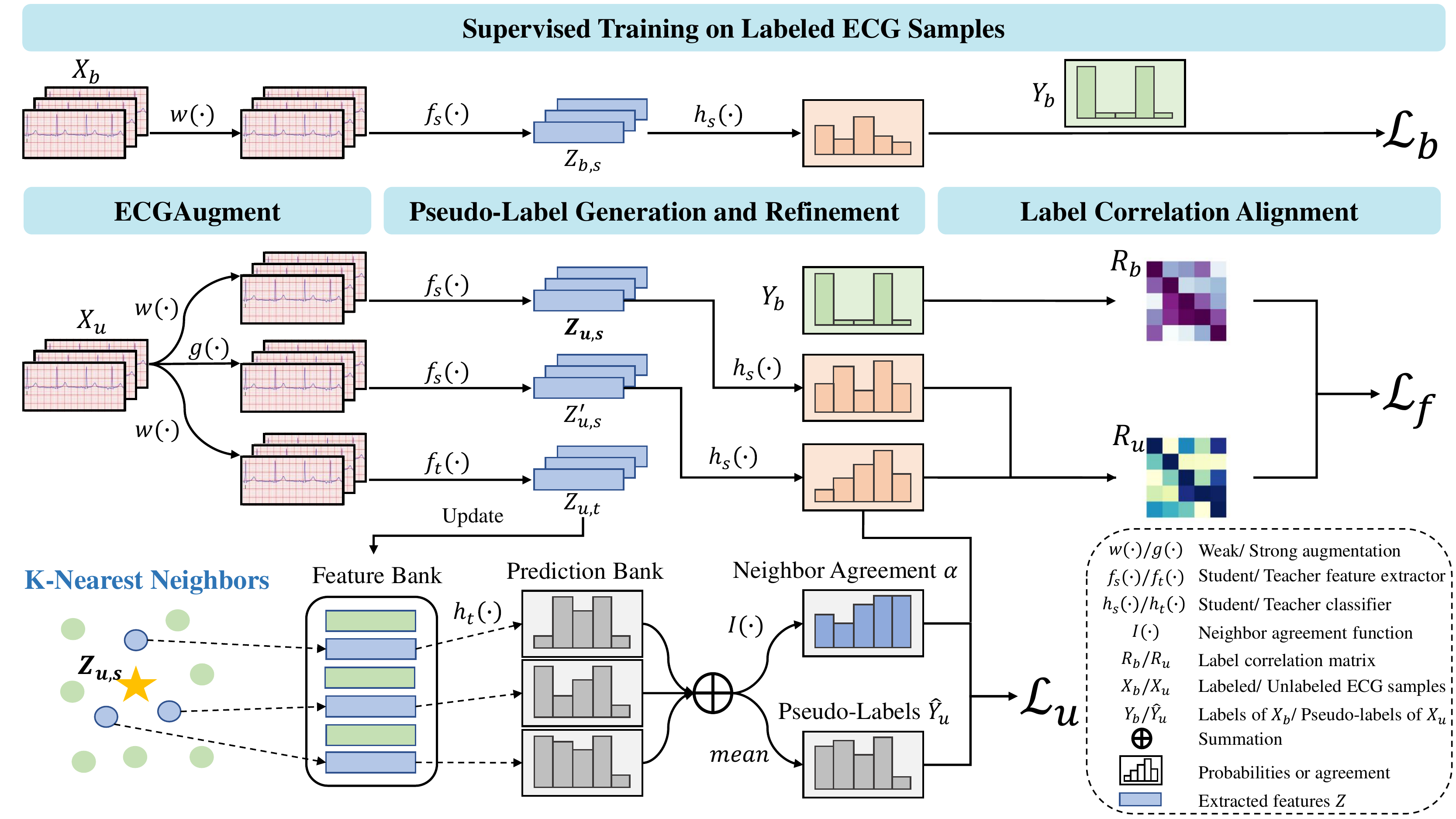}
\end{center}
\caption{Overall schematics of ECGMatch. It consists of three losses and four parts. \textbf{1) Supervised training on the labeled ECG samples}: the student network $M_s=\{f_s(\cdot),h_s(\cdot)\}$ outputs CVDs predictions for the labeled samples $X_b$ and computes the supervised loss $\mathcal{L}_b$ in Eq.\ref{Eq:supervisedlabelloss}. \textbf{2) ECGAugment for the unlabeled samples}: apply weak and strong augmentations to the unlabeled samples $X_u$. \textbf{3) Pseudo-label generation and refinement}: generate pseudo-labels for the unlabeled samples using two memory banks maintained by the teacher network $M_t=\{f_t(\cdot),h_t(\cdot)\}$; refine the raw pseudo-labels based on a neighbor agreement function $I(\cdot)$; computes the unsupervised loss $\mathcal{L}_u$ defined in Eq.\ref{Eq:refinepseudolabelloss}. \textbf{4) Label correlation alignment}: estimate the label correlation matrices for the labeled and unlabeled samples and compute the loss $\mathcal{L}_f$ in Eq.\ref{Eq:Frobenius}.} 
\label{fig:flowchart}
\end{figure*}

\section{Methodology}
\label{sec:methodology}
\subsection{Overview}
In semi-supervised learning for multi-label CVDs prediction, the training ECG data is divided into the labeled and unlabeled sets, given as $D_B=\{X_b,Y_b\}=\{x_b^i,y_b^i\}_{i=1}^{N_B}$ and $D_U=\{X_u,-\}=\{x_u^i,-\}_{i=1}^{N_U}$. $N_B$ and $N_U$ are the number of samples in $D_B$ and $D_U$. $X_b$ contains the labeled 12-leads ECG recordings, and $Y_b$ represents the corresponding multi-label ground-truth. Specifically, the $c$-th dimension in $y_b^i$ contains the ground-truth of category $c$, $y_b^{i,c}\in \{0,1\}$. Hence, a given ECG recording might belong to multiple categories simultaneously. To clarify the narrative, the frequently used notations are summarized in Table \ref{tab:frequentNotations}. As shown in Fig. \ref{fig:flowchart}, the proposed ECGMatch includes three modules: ECGAugment module, pseudo-label generation and refinement module, and label correlation alignment module. \textbf{In the ECGAugment module}, motivated by the weak-strong augmentation method\cite{sohn2020fixmatch}, we design a novel augmentation pipeline by investigating the intrinsic characteristics of the ECG signals, named as ECGAugment. \textbf{In the pseudo-label generation and refinement module}, we introduce a knowledge distillation method for pseudo-label generation. Then, we propose a neighbor agreement modeling method to compute the importance score for the pseudo labels, which can alleviate the negative effect of the inaccurate pseudo-labels. \textbf{In the label correlation alignment module}, we propose to align the label correlation matrices computed on the labeled data and unlabeled data by Frobenius norm regularization, which enables the model to capture the label dependency between different CVDs. More details about the proposed ECGMatch are presented below.
\begin{table}
\begin{center}
\caption{Frequently used notations and descriptions.}
\label{tab:frequentNotations}
\begin{tabular}{cc}
\toprule
 Notation             & Description   \\
\midrule
 $B$ & batch size\\
 $C$ & the number of CVDs category\\
 $R$ & label correlation matrix\\
 $K$ & the number of nearest neighbor\\
 $\alpha$ & neighbor agreement\\
 $D_B    \backslash D_U $ & labeled\textbackslash{}unlabeled datasets\\
 $Y_b \backslash \hat{Y}_u$ & true\textbackslash{}pseudo labels\\
 $X_b \backslash X_u$ & labeled\textbackslash{}unlabeled ECG samples\\
 $Z\backslash P$ & feature\textbackslash{}prediction banks\\
 $M_s \backslash M_t$ & student\textbackslash{}teacher networks\\
 $w(\cdot) \backslash g(\cdot)$ & weak\textbackslash{}strong augmentations\\
  $f(\cdot) \backslash h(\cdot)$ & feature extractor\textbackslash{}multi-label classifier\\
\bottomrule
\end{tabular}
\end{center}
\end{table}

\subsection{ECGAugment}
One critical method to tackle the label scarcity problem is efficient data augmentation\cite{UDA2020}. Although ECG augmentation methods have been well investigated in previous studies\cite{raghu2022data,zhang2022semi}, how to properly define a weak and strong augmentation pipeline for ECG-based semi-supervised learning is still challenging. Hence, we propose a novel augmentation pipeline for the ECG signals by leveraging their characteristic, termed as ECGAugment. Specifically, 'weak' augmentation $w(\cdot)$ is defined by randomly choosing one transformation to augment a 12-lead ECG signal $x\in \mathbb{R}^{12 \times L}$, where $L$ is the length of $x$. 1. \textbf{Signal Dropout}: we randomly set the ECG signal values within a random time window to zero. Its length and location are randomly generated from uniform distributions. This transformation enables the model to handle weak signals caused by bad contact of ECG electrodes\cite{zihlmann2017convolutional}. 2. \textbf{Temporal Flipping}: motivated by previous studies\cite{kiyasseh2021clocs,nonaka2020electrocardiogram}, we flip the original ECG signal along the temporal axis, which means the signal is read in reverse. 
3. \textbf{Channel Reorganization}: Each row of $x$ represents the ECG signal recorded at one lead (channel). Hence, we randomly change the order of the row vectors in the signal matrix $x$ to shuffle its channel organization. 4. \textbf{Random Noise}: inspired by the noise contamination technique in ECG-based contrastive learning and adversarial learning\cite{kiyasseh2021clocs,han2020deep}, we add a Gaussian noise $\epsilon \sim \mathcal{N}(0, \sigma)$ to the original signal $x$. 

Motivated by the RandAugment technique for image augmentation\cite{cubuk2020randaugment}, we define the 'strong' augmentation $g(\cdot)$ by randomly selecting $T\leq4$ transformations to perturb the input signal $x$. Specifically, a transformation queue is randomly generated and transformations within the queue are applied one after another. For a random queue $\{2,1,3\}$, we successively apply Temporal Flipping, Signal Dropout, and Channel Reorganization to the input signal. Compared with traditional sequential perturbations which fix the number and the order of transformations\cite{kiyasseh2021clocs,oliveira2022generalizable}, the proposed method dramatically increases the diversity of the augmented samples by introducing extra randomness, which greatly increases the model performance\cite{UDA2020,cubuk2020randaugment}.
\subsection{Pseudo-Label Generation for Multi-Label Learning}
\label{sec:generation}
The key to robust semi-supervised learning is accurate pseudo-label generation, which has been demonstrated by many previous studies\cite{berthelot2019mixmatch,sohn2020fixmatch,zhang2021flexmatch,wang2022freematch,chen2022contrastive,gao2022visual}. However, previous studies mainly consider a single-label condition, where each sample belongs to one category only. In contrast, we focus on a multi-label condition in this study, where each sample belongs to multiple categories simultaneously. Here, we generate the pseudo-labels using a knowledge distillation method. Specifically, we introduce a teacher model $M_t=\{f_t(\cdot), h_t(\cdot)\}$ and a student model $M_s=\{f_s(\cdot), h_s(\cdot)\}$, where $f(\cdot)$ is a feature extractor and $h(\cdot)$ is a multi-label classifier. As shown in Fig. \ref{fig:flowchart}, we first apply the weak augmentation $w(\cdot)$ and the strong augmentation $g(\cdot)$ to the unlabeled ECG recordings $x_u$, respectively. The teacher model extracts deep features $z_{u,t}=f_t(w(x_u))$ from the weak-augmented signals $w(x_u)$ and outputs the corresponding predictions $p_{u,t}=sigmoid(h_t(z_{u,t}))$. Then we store them in two memory banks (feature bank $Z=\{z_{u,t}^n\}_{n=1}^{N_U}$ and prediction bank $P=\{p_{u,t}^n\}_{n=1}^{N_U}$), $N_U$ is the number of samples in ${D}_{U}$. Note that $Z$ and $P$ are updated on the fly with the current mini-batch. In this study, $p_{u,t}^n=[p_{u,t}^{n,1},..., p_{u,t}^{n,c}]$ is a $C$ dimensional vector where the $c$-th element represents the prediction of class $c$, $p_{u,t}^{n,c} \in [0,1]$. During training, the student model extracts a feature vector $z_{u,s}^i=f_s(w(x_u^i))$ from a given unlabeled sample $x_u^i$ and assigns a pseudo-label ($\hat{y}^i_u$) for it using a widely used soft voting method\cite{chen2022contrastive}. Specifically, $\hat{y}^i_u$ is computed by integrating the predictions of its K-Nearest neighbors $\{z_{u,t}^k\}_{k=1}^{K}$ in the feature bank $Z$, given as
\begin{equation}
\label{Eq:pseudolabel}
\hat{y}^i_u=\frac{1}{K}\sum_{k=1}^{K}p_{u,t}^k,
\end{equation}
where $p_{u,t}^k$ is the prediction of $z_{u,t}^k$, which is the $k$-th nearest neighbor of the feature vector $z_{u,s}^i$ in the feature bank $Z$, $K$ is the number of neighbors. $\{p_{u,t}^k\}_{k=1}^{K}$ is acquired by visiting the prediction bank $P$. To conduct the knowledge distillation process, we minimized the binary cross entropy loss between the prediction of the student model and the pseudo-label $\hat{y}^i_u$ given by the prediction bank. Motivated by the weak-strong consistency regularization method\cite{sohn2020fixmatch}, we apply a strong augmentation $g(\cdot)$ to the unlabeled sample $x_u^i$ and compute the corresponding student prediction by $q_{u,s}^i=sigmoid(h_s(z_{u,s}^{'})),z_{u,s}^{'}=f_s(g(x_u^i))$. Then we compute the binary cross entropy loss between the pseudo-labels and the student predictions of the unlabeled samples, defined as 
\begin{equation}
\label{Eq:pseudolabelloss}
\mathcal{L}_u=-\frac{1}{B_uC}\sum^{B_u}_{i=1}\sum^{C}_{c=1}(1-\hat{y}^{i,c}_u)\log(1-q_{u,s}^{i,c})+\hat{y}^{i,c}_u\log q_{u,s}^{i,c},
\end{equation}
where $C$ is the number of categories in the dataset, and $B_u$ is the number of unlabeled samples in the current mini-batch. Using the ground truth of the labeled samples in the mini-batch, we compute the supervised binary cross-entropy loss, defined as
\begin{equation}
\label{Eq:supervisedlabelloss}
\mathcal{L}_b=-\frac{1}{BC}\sum^{B}_{i=1}\sum^{C}_{c=1}(1-y^{i,c}_b)\log(1-p_{b,s}^{i,c})+y^{i,c}_b\log p_{b,s}^{i,c},
\end{equation}
where $B$ is the number of labeled samples in the current mini-batch, $p_{b,s}^{i,c}=sigmoid(h_s(f_s(w(x_b^i))))$ is the prediction outputed by the student model and $y^{i,c}_b \in\{0,1\}$ is the corresponding ground truth. Combing Eq. \ref{Eq:pseudolabelloss} and Eq. \ref{Eq:supervisedlabelloss}, we compute the overall loss for semi-supervised multi-label classification, defined as 
\begin{equation}
\label{Eq:semiloss}
\mathcal{L}=\mathcal{L}_b + \lambda\mathcal{L}_u,
\end{equation}
where $\lambda$ is a hyper-parameter controlling the weight of $\mathcal{L}_u$. Before pseudo-label generation, the teacher model $M_t$ is pre-trained on the labeled dataset $D_B$ using the Eq.\ref{Eq:supervisedlabelloss}. Then in the knowledge distillation process, the student model $M_s$ is updated by stochastic gradient descent to minimize Eq. \ref{Eq:semiloss}. To stabilize the maintained feature bank $Z$ and the prediction bank $P$, the parameters $\theta_t$ of the teacher model $M_t$ are updated by the momentum moving average of the parameters $\theta_s$ of the student model\cite{he2020momentum,chen2022contrastive}, defined as
\begin{equation}
\label{Eq:ema}
\theta_t=m\theta_t+(1-m)\theta_s,
\end{equation}
where $m$ is a momentum hyper-parameter. 
\subsection{Pseudo-Label Refinement based on Neighbor Agreement Modeling}
\label{sec:refinement}
Inaccurate pseudo-labels can hurt the model performance in semi-supervised learning. Consequently, the generated pseudo-labels $\hat{y}_u$ should be further refined to avoid this problem. Previous studies\cite{sohn2020fixmatch,zhang2021flexmatch} utilized fixed or dynamic thresholds to remove the pseudo-labels with low confidence. However, it is difficult to set up separate optimized thresholds for different categories in multi-label classification. Moreover, designing update strategies for dynamic thresholds needs tremendous hyper-parameters\cite{chen2023softmatch,wang2022freematch}. Hence, we proposed a novel pseudo-label refinement method based on \emph{neighbors agreement modeling} (NAM). It refines the raw pseudo-labels $\hat{y}_u$ by computing their neighbor agreement based on a neighbor agreement function $I(\cdot)$ and then adjusts their importance in the loss propagation. \textcolor{black}{Compared with traditional threshold-based refinement method\cite{sohn2020fixmatch,zhang2021flexmatch,huang2022percentmatch}, NAM replaces the threshold control process with an importance weighting process, which is more hyperparameter-efficient in semi-supervised multi-label classification.}

Recall that we have generated the raw pseudo-label $\hat{y}^i_u$ for the unlabeled sample $x_u^i$ by averaging the prediction $p_{u,t}^k$ of its K-Nearest neighbors (Eq. \ref{Eq:pseudolabel}). Here, we sum up the neighbors' predictions $p_{u,t}^{k,c}\in [0,1]$ and apply a neighbor agreement function $I(\cdot)$ to compute the neighbor agreement of the pseudo-label $\hat{y}^i_u$ on the $c$-th category.
\begin{equation}
\label{Eq:agreement}
\alpha^{i,c}_u=I(\sum_{k=1}^{K}p_{u,t}^{k,c})=\left|\frac{2}{K}\sum_{k=1}^{K}p_{u,t}^{k,c}-1\right|,
\end{equation}
where $K$ is the number of nearest neighbors. $\alpha^{i,c}_u \in [0,1]$ is the neighbor agreement which can also be regarded as the model confidence on the pseudo-label $\hat{y}^i_u$. Combing the Eq. \ref{Eq:agreement} and Eq. \ref{Eq:pseudolabelloss}, we can rewrite the unsupervised binary cross entropy loss as
\begin{equation}
\label{Eq:refinepseudolabelloss}
\mathcal{L}_u=-\frac{1}{B_uC}\sum^{B_u}_{i=1}\sum^{C}_{c=1}\alpha^{i,c}_u[(1-\hat{y}^{i,c}_u)\log(1-q_{u,s}^{i,c})+\hat{y}^{i,c}_u\log q_{u,s}^{i,c}],
\end{equation}
where $\alpha^{i,c}_u$ controls the weight of the $\hat{y}^{i,c}_u$ in loss computation. Specifically, the Eq. \ref{Eq:agreement} allocates high weights ($\alpha^{i,c}_u \approx 1$ ) to the pseudo-labels with high agreement on neighbors' prediction ($\sum_{k=1}^{K}p_{u,t}^{k,c} \approx K$ or 0). Note that the label refinement process of the NAM module is only related to one hyperparameter $K$, which is the number of nearest neighbors. On the contrary, previous threshold-based methods need to set up fixed or dynamic thresholds for $C$ independent categories in multi-label classification, which is less efficient in hyper-parameters grid searching than the proposed NAM module. \textcolor{black}{In addition, threshold-based methods typically discard  pseudo-labels whose confidences are lower than the pre-defined thresholds. The selection of the thresholds is sensitive to the class distribution in training datasets\cite{berthelot2019remixmatch,sohn2020fixmatch}, which can result in suboptimal generalization performance when applied to the unseen dataset with a different class distribution.  In contrast, the proposed NAM employs a soft method that adjusts the importance of the generated pseudo-labels rather than directly rejecting them. It is less sensitive to the class distribution in the training data compared with the threshold-based methods\cite{chen2023softmatch} and can enhance the model performance on the unseen dataset.}
\subsection{Label Correlation Alignment}
\label{sec:alignment}
The co-occurrence of CVDs leads to a strong relationship between different categories, which should be considered to achieve better prediction performance in multi-label classification\cite{zhang2007ml,liu2017easy}. Previous studies focused on the label dependency within the labeled samples and utilized the semantic relationship between different categories to guide the model training\cite{yeh2017learning,wang2021semi,liang2022multi}. However, it is hard to define the relationship among various CVDs without sufficient prior knowledge. On the other hand, ignoring the label dependency within the unlabeled sample results in unnecessary information waste. Hence, we propose jointly capturing the label dependency within the labeled and unlabeled samples by computing two label correlation matrices ($R_b$ and $R_u$). In practice, $R_b$ is calculated using the labeled samples while $R_u$ is estimated using the unlabeled samples. The computation process does not need extra prior information like the word-embedding correlation between different labels\cite{wang2021semi}. Then we minimize the discrepancy between the $R_b$ and $R_u$ to align the label dependencies computed by the labeled and unlabeled samples, which enhances the model performance in multi-label classification.

Firstly, we introduce how to compute the label correlation matrix $R_b$ based on the labeled sample set $D_B=\{X_b, Y_b\}$. $Y_b=[y_b^1;y_b^2;...;y_b^{N_B}]$ is a $N_B \times C$ label matrix, where $C$ is the number of categories. The label correlation $\hat{r}_{c_1,c_2}\in[0,1]$ between classes $c_1$ and  $c_2$ can be estimated by the similarity between the label sequences ($y_{c_1}$,$y_{c_2}$) on the two classes, where $y_{c_1}=[y_b^{1,c_1};y_b^{2,c_1},...,y_b^{N_B,c_1}]$ and $y_{c_2}=[y_b^{1,c_2};y_b^{2,c_2},...,y_b^{N_B,c_2}]$. We find that cosine similarity is more efficient for label correlation analysis than other metrics such as the Pearson coefficient. As shown in Eq.\ref{Eq:cosine_similarity}, with binarized labels $y_{c_1}$ and $y_{c_2}$, it estimates the conditional probabilities between the classes $c_1$ and $c_2$ without being influenced by the class distributions of different datasets. The proof of Eq.\ref{Eq:cosine_similarity} and detailed analysis of different similarity metrics can be found in Appendix \ref{sec:sim}. Based on the cosine similarity, the correlation $\hat{r}_{c_1,c_2}$ is computed as
\begin{equation}
\label{Eq:cosine_similarity}
\hat{r}_{c_1,c_2}=\frac{y_{c_1}^Ty_{c_2}}{\left\|y_{c_1}\right\|\left\|y_{c_2}\right\|}=\sqrt{P(c_1=1|c_2=1)P(c_2=1|c_1=1)},
\end{equation}
And the label correlation matrix $R_b$ can be computed by
\begin{equation}
\label{Eq:correlation}
R_b=\begin{bmatrix} 
	\hat{r}_{1,1} &\hat{r}_{1,2}&\cdots&\hat{r}_{1,C} \\
	\hat{r}_{2,1} &\hat{r}_{2,2}&\cdots&\hat{r}_{2,C}\\
	\vdots&\vdots& \ddots&\vdots \\
	\hat{r}_{C,1}&\hat{r}_{C,2}&\cdots&\hat{r}_{C,C}
	\end{bmatrix}=N(Y)^TN(Y),
\end{equation}
where $N(Y)$ is a normalization function which normalizes the column vectors of $Y$ to unit vectors. For the unlabeled sample, we estimate the label correlation matrix $R_u$ using the model prediction $P_u$ output by the student network $M_s$. To improve the robustness of the estimated $R_u$, we simultaneously use the strongly-augmented and weakly-augmented samples to increase the sample size for computation. Hence, $R_u$ is estimated as
\begin{equation}
\label{Eq:correlation_u}
R_u=N(P_u)^TN(P_u), P_u=[q_{u,s}^1;p_{u,s}^1;...;q_{u,s}^{B_u};p_{u,s}^{B_u}],
\end{equation}
where $P_u$ is a $2B_u\times C$ matrix containing the student predictions of the strongly and weakly augmented unlabeled samples ($g(x_u)$ and $w(x_u)$), and $B_u$ is the number of unlabeled samples in the current mini-batch. Specifically, $q_{u,s}^i=sigmoid(h_s(f_s(g(x_u^i))))$ and $p_{u,s}^i=sigmoid(h_s(f_s(w(x_u^i))))$. The label correlation matrices represent the dependency and the semantic relationship between different CVDs, which should be consistent across the labeled and unlabeled data. Consequently, we minimize the discrepancy between the $R_u$ and $R_b$ using Frobenius norm regularization, defined as
\begin{equation}
\label{Eq:Frobenius}
\mathcal{L}_f=\left\|R_b-R_u\right\|_F,
\end{equation}
where $\left\|\cdot\right\|_F$ represents the Frobenius norm of a given matrix. Finally, we formulate the final loss of the proposed ECGMatch by combing the the supervised multi-label classification loss (Eq. \ref{Eq:supervisedlabelloss}), the importance weighted unsupervised multi-label classification loss (Eq. \ref{Eq:refinepseudolabelloss}) and the label correlation alignment loss (Eq. \ref{Eq:Frobenius})
\begin{equation}
\label{Eq:finalloss}
\mathcal{L}=\mathcal{L}_b + \lambda_u\mathcal{L}_u+\lambda_f\mathcal{L}_f,
\end{equation}
where $\lambda_u$ and $\lambda_f$ are two hyper-parameters controlling the importance of different objective functions. We present the complete algorithm for ECGMatch in algorithm \ref{alg::conjugateGradient}.
\begin{algorithm*}[h]
  \caption{ECGMatch algorithm}
  \label{alg::conjugateGradient}
  \begin{algorithmic}[1]
    \Require
\renewcommand{\algorithmicrequire}{\textbf{}}
    \Require - Label dataset $D_B=\{X_b,Y_b\}=\{x_b^i,y_b^i\}_{i=1}^{N_B}$ and unlabeled dataset $D_U=\{X_u,-\}=\{x_u^i,-\}_{i=1}^{N_U}$;
    \Require - Student model $M_s$ and teacher model $M_t$; Feature bank $Z$ and prediction bank $P$; Batch size $B$
    \Ensure Trained student model $M_s$; 
    \State pretrain the teacher model using Eq. \ref{Eq:supervisedlabelloss} and $D_B$; compute the label correlation matrix $R_b$ using Eq.\ref{Eq:correlation};
    \For {$1$ to $Epoch$} 
         \For {1 to $iteration$} \textcolor{gray}{\qquad\qquad\qquad\qquad\qquad\qquad\qquad\qquad\qquad\qquad\qquad\qquad//$iteration=\frac{N_B}{B}$ }
         \Require \textcolor{gray}{\#\# Mini-batch sampling and ECGAugment \#\#}
          \State sample labeled data $\{x_b,y_b\}$ from $D_B$; 
          \State sample unlabeled data $\{x_u,-\}$ from $D_U$;
          \State apply ECGAugment to $x_b$ and $x_u$;
          \State compute the supervised loss $\mathcal{L}_b$ using $\{x_b,y_b\}$ and Eq.\ref{Eq:supervisedlabelloss};
          \Require \textcolor{gray}{\#\# Pseudo-label generation \#\#}
          \State update the feature and prediction banks using $x_u$ and $M_t$;
          \State generate pseudo-labels $\hat{y}_u$ for $x_u$ using Eq.\ref{Eq:pseudolabel};
          \Require \textcolor{gray}{\#\# Pseudo-label refinement \#\#}
          \State compute the neighbor agreement $w_u$ of $\hat{y}_u$ using Eq.\ref{Eq:agreement};
          \State compute the unsupervised loss $\mathcal{L}_u$ using Eq.\ref{Eq:refinepseudolabelloss} 
          \Require \textcolor{gray}{\#\# Label correlation alignment \#\#}
          \State compute the label correlation matrix $R_u$ using Eq.\ref{Eq:correlation_u};
          \State compute the loss $\mathcal{L}_f$ using Eq.\ref{Eq:Frobenius}; compute the final loss $\mathcal{L}$ using Eq.\ref{Eq:finalloss};
          \State update network $M_s$ by minimizing $\mathcal{L}$ by stochastic gradient descent; update network $M_t$ using Eq.\ref{Eq:ema}.
    \EndFor
    \State apply an early-stop strategy to avoid overfitting;
\EndFor
\end{algorithmic}
\end{algorithm*}

\section{Experiments and Datasets}
\label{sec:experiment}
\subsection{Public ECG Databases}
To evaluate the performance of the proposed ECGMatch model, we conduct experiments on four well-known public databases released on the PhysioNet website: The Georgia 12-lead ECG Challenge (G12EC) Database\cite{alday2020classification}, the Physikalisch-Technische Bundesanstalt (PTB-XL) database\cite{wagner2020ptb}, the Chapman-Shaoxing databases\cite{zheng202012} and the Ningbo databases\cite{zheng2020optimal}. The G12EC database contains 10,344 available ECG recordings, each lasting between 5 and 10 seconds long with a sampling frequency of 500 Hz. The PTB-XL database contains 22,353 available ECG recordings and each recording is around 10 seconds long at a sampling frequency of 500 Hz. The Chapman-Shaoxing database consists of ECG recordings from 10,646 subjects, while the Ningbo database contains 40258 ECG recordings, both of which were sampled at a frequency of 500 Hz. Unfortunately, the aforementioned databases employed significantly different label annotation schemes and contained different kinds of CVDs, which led to a substantial category gap across databases. As a detailed discussion about the category gap problem is beyond the scope of our study, we simply addressed this issue by using a consistent label annotation scheme to re-annotate the databases. In summary, we re-annotate the ECG signals from the datasets by categorizing them into five classes (Abnormal Rhythms, ST/T Abnormalities, Conduction Disturbance, Other Abnormalities, and Normal Signals). Note that the ECG signals might belong to two or more categories simultaneously. Details about the annotation scheme can be found in Appendix \ref{sec:dataset}. To preprocess the signals, we first normalize the length of the raw signals into 6144 samples in the time domain by zero-padding. Next, we apply a bandpass filter (1.0-47.0 Hz) to eliminate noise components within the raw ECG recordings. Finally, the signals are normalized using z-score normalization.   
\vspace{-0.5em}
\subsection{Implementation Details}
In our implementation, we use the Attention-based Convolutional Neural Network\cite{AIMTbackbone} as the feature extractor $f(\cdot)$ in Fig.\ref{fig:flowchart}, where the dimension of the output feature $z$ is 128. The classifier $h(\cdot)$ is designed as 128 neurons (input layer)-128 neurons (hidden layer 1)-5 neurons (output layer)-Sigmoid activation. The teacher network $M_t=\{f_t,h_t\}$ is pre-trained on the labeled sample set $D_B$, and the parameters of the student network $M_s=\{f_s,h_s\}$ are initialized with those of $M_t$. In the semi-supervised training process, the parameters of $M_t$ are updated by Eq.\ref{Eq:ema}, with a momentum of 0.999. We use the standard stochastic gradient descent (SGD) optimizer with a momentum of 0.9 for parameter optimization. The initial learning rate is set to 3e-2 with an exponential learning rate decay schedule as $\eta=\frac{\eta_0}{(1+\gamma e/E)^{-p}}$, where $\eta_0$ is the initial learning rate, $e$ is the current training step and $E=5000$ is the max training step. In each mini-batch, the number of labeled samples $B$ is 64 and the number of unlabeled samples $B_u$ is 448. The weights $\lambda_u$ and $\lambda_f$ in Eq. \ref{Eq:finalloss} are searched within the range of $[0,1.6]$ with a step of 0.4. 

\subsection{Experimental Protocols for Model Evaluation}

To assess the robustness of the proposed model on multi-label CVDs classification, we propose three distinct protocols for model evaluation when taking into account various clinical applications. \textbf{1) Within-dataset protocol.} For model training and evaluation, the training, validation, and testing data are randomly sampled from one dataset in a ratio of 0.8 : 0.1 : 0.1. Then, we split the training data into labeled and unlabeled data in a ratio of 0.05 : 0.95. Finally, the average performance and standard deviations of four datasets are computed across three random seeds. \textbf{2) Mix-dataset protocols.} In this scenario, we randomly sample the training, validation, and testing data from four datasets simultaneously in a ratio of 0.8 : 0.1 : 0.1. The training data is split into labeled and unlabeled data in a ratio of 0.01 : 0.99. The average performance and standard deviations are calculated across three random seeds. This protocol considers a multi-center setting, where the training data contains samples from different datasets (centers).  \textbf{3) Cross-dataset protocols.} To evaluate the model performance on the unseen testing dataset (s), we use three datasets for model training and validation and reserve the remaining one for testing. For example, we can reserve the G12EC dataset as the unseen test set and sample the training (90\%) and validation data (10\%) from the remaining three datasets (PTB-XL, Chapman, Ningbo). Only 1\% of the training data is labeled, while the remaining 99\% is unlabeled.  We repeat the evaluation process until each dataset is used once as the unseen test set and report the average performance and standard deviations across three random seeds.  This protocol serves as an external validation of the proposed model, which evaluates the model's generalization ability across different independent datasets. 

We evaluate the performance of various models using multiple multi-label metrics including ranking loss, hamming loss, coverage, mean average precision (MAP), macro AUC and marco-$G_{beta}$. It is important to note that lower values of  ranking loss, hamming loss and coverage indicate better performance, while lower values of  MAP, macro AUC and marco-$G_{beta}$ score mean worse performance. More details about these metrics can be found in \cite{zhang2013review}. The following section presents a comparison between the proposed ECGMatch and the existing literature based on the three experimental protocols and six evaluation metrics mentioned above. As there is limited research on the application of SSL for ECG-based multi-label classification, we replicated several state-of-the-art (SOTA) models that were originally implemented for image or text classification: MixMatch\cite{berthelot2019mixmatch}, FixMatch\cite{sohn2020fixmatch}, FlexMatch\cite{zhang2021flexmatch}, DST\cite{chen2022debiased}, PercentMatch\cite{huang2022percentmatch}, SoftMatch\cite{chen2023softmatch}, UPS\cite{rizve2021defense}. We ensure consistency across all compared models by employing identical backbones and augmentation strategies (ECGAugment). We also use the same set of common hyper-parameters, such as learning rate and batch size. For the model-specific parameters such as the sharpen-temperature in FixMatch\cite{sohn2020fixmatch}, we utilize the optimal settings recommended by the referenced studies.

\section{Results and Discussion}
\label{sec:discussion} 
\subsection{Comparisons with State-of-the-Art Methods}
The performance of different models on different protocols is presented in Table \ref{tab:Withincompare}, Table \ref{tab:Crosscompare} and Table \ref{tab:mixcompare}.  The results show that ECGMatch achieves the leading performance in all experimental protocols, which demonstrates its superiority. On the one hand, the averaged performance of EEGMatch is better than the threshold-based SOTA models, such as FixMatch\cite{sohn2020fixmatch}, FlexMatch\cite{zhang2021flexmatch}, DST\cite{chen2022debiased}, especially when the test data comes from an unseen dataset. Specifically, the performance difference between the ECGMatch and the other models in the cross-dataset protocol is more distinct than that in the within-dataset and mix-dataset protocols. This phenomenon suggests that the NAM module is more efficient for pseudo-label refinement in multi-label classification than fixed or dynamic threshold strategies, especially on unseen datasets. On the other hand, we also notice that the ECGMatch achieves better performance than PercentMatch\cite{huang2022percentmatch} and UPS\cite{rizve2021defense}, which are the latest models designed for semi-supervised multi-label learning. This observation indicates that capturing label relationships within the labeled and unlabeled samples benefits multi-label classification, while this property is ignored in the above two competitors. More details about the contributions of each component are listed in the next sub-section. In summary, the remarkable improvements in different protocols demonstrate the potential of ECGMatch to be implemented in various clinical applications.

\begin{table*}[h]
\setlength{\tabcolsep}{0.86em}
\begin{center}
\caption{Comparison results between ECGMatch and the state-of-the-art models using the within-dataset protocol. The mean performance and standard deviations on four databases are shown across three seeds.}
\label{tab:Withincompare}
\scalebox{1}{
\color{black}
\begin{tabular*}{\linewidth}{@{}lccccccccc@{}}
\toprule
Methods & MixMatch\cite{berthelot2019mixmatch} & FixMatch\cite{sohn2020fixmatch} & FlexMatch\cite{zhang2021flexmatch} & DST\cite{chen2022debiased} & PerMatch\cite{huang2022percentmatch} & SoftMatch\cite{chen2023softmatch} & UPS\cite{rizve2021defense} & \textbf{ECGMatch} \\
\midrule
\multicolumn{9}{c}{\textbf{Ranking loss} (The smaller, the better)}\\
\midrule
G12EC & 0.349$\pm$0.033&0.217$\pm$0.041&0.160$\pm$0.009&0.189$\pm$0.019&0.167$\pm$0.006&0.199$\pm$0.045&0.177$\pm$0.016&\textbf{0.140$\pm$0.006}\\
PTB-XL & 0.345$\pm$0.004&0.170$\pm$0.010&0.146$\pm$0.004&0.279$\pm$0.148&0.200$\pm$0.065&0.158$\pm$0.016&0.156$\pm$0.001&\textbf{0.134$\pm$0.003}\\
Ningbo & 0.178$\pm$0.014&0.106$\pm$0.048&0.153$\pm$0.024&0.082$\pm$0.020&0.200$\pm$0.020&0.197$\pm$0.047&0.172$\pm$0.082&\textbf{0.045$\pm$0.002}\\
Chapman & 0.214$\pm$0.027&0.103$\pm$0.051&0.088$\pm$0.014&0.146$\pm$0.099&0.080$\pm$0.008&0.122$\pm$0.014&0.075$\pm$0.008&\textbf{0.052$\pm$0.002}\\
\midrule
\multicolumn{9}{c}{\textbf{Hamming loss} (The smaller, the better)}\\
\midrule
G12EC & 0.538$\pm$0.032&0.306$\pm$0.004&0.303$\pm$0.004&0.330$\pm$0.013&0.321$\pm$0.009&0.311$\pm$0.009&0.294$\pm$0.008&\textbf{0.278$\pm$0.008}\\
PTB-XL & 0.439$\pm$0.018&0.257$\pm$0.005&0.265$\pm$0.013&0.407$\pm$0.207&0.251$\pm$0.005&0.265$\pm$0.021&0.255$\pm$0.006&\textbf{0.233$\pm$0.009}\\
Ningbo & 0.421$\pm$0.094&0.148$\pm$0.015&0.139$\pm$0.005&0.134$\pm$0.004&0.136$\pm$0.008&0.138$\pm$0.005&0.136$\pm$0.003&\textbf{0.122$\pm$0.001}\\
Chapman & 0.423$\pm$0.070&0.168$\pm$0.009&0.189$\pm$0.008&0.180$\pm$0.009&0.182$\pm$0.006&0.196$\pm$0.003&0.167$\pm$0.010&\textbf{0.139$\pm$0.002}\\
\midrule
\multicolumn{9}{c}{\textbf{Coverage} (The smaller, the better)}\\
\midrule
G12EC & 2.990$\pm$0.115&2.471$\pm$0.143&2.255$\pm$0.037&2.369$\pm$0.085&2.281$\pm$0.031&2.395$\pm$0.163&2.325$\pm$0.064&\textbf{2.173$\pm$0.027}\\
PTB-XL & 2.728$\pm$0.007&2.065$\pm$0.028&1.971$\pm$0.015&2.481$\pm$0.555&2.187$\pm$0.261&2.007$\pm$0.061&2.016$\pm$0.009&\textbf{1.922$\pm$0.015}\\
Ningbo & 2.317$\pm$0.054&1.978$\pm$0.200&2.164$\pm$0.094&1.880$\pm$0.080&2.346$\pm$0.086&2.347$\pm$0.186&2.241$\pm$0.329&\textbf{1.724$\pm$0.010}\\
Chapman & 2.466$\pm$0.128&1.981$\pm$0.190&1.912$\pm$0.038&2.121$\pm$0.362&1.888$\pm$0.038&2.040$\pm$0.041&1.859$\pm$0.038&\textbf{1.761$\pm$0.021}\\
\midrule
\multicolumn{9}{c}{\textbf{MAP} (The greater, the better)}\\
\midrule
G12EC & 0.479$\pm$0.012&0.703$\pm$0.008&0.719$\pm$0.008&0.690$\pm$0.016&0.711$\pm$0.007&0.717$\pm$0.010&0.719$\pm$0.011&\textbf{0.742$\pm$0.005}\\
PTB-XL & 0.546$\pm$0.023&0.737$\pm$0.013&0.738$\pm$0.012&0.739$\pm$0.005&0.737$\pm$0.014&0.730$\pm$0.012&0.740$\pm$0.014&\textbf{0.748$\pm$0.009}\\
Ningbo & 0.484$\pm$0.059&0.796$\pm$0.006&0.793$\pm$0.005&0.791$\pm$0.006&0.786$\pm$0.005&0.794$\pm$0.002&0.797$\pm$0.005&\textbf{0.808$\pm$0.001}\\
Chapman & 0.500$\pm$0.073&0.730$\pm$0.005&0.732$\pm$0.005&0.721$\pm$0.015&0.736$\pm$0.005&0.736$\pm$0.004&0.737$\pm$0.012&\textbf{0.775$\pm$0.014}\\
\midrule
\multicolumn{9}{c}{\textbf{Marco AUC} (The greater, the better)}\\
\midrule
G12EC & 0.668$\pm$0.023&0.841$\pm$0.004&0.850$\pm$0.004&0.833$\pm$0.014&0.843$\pm$0.005&0.846$\pm$0.004&0.848$\pm$0.005&\textbf{0.854$\pm$0.003}\\
PTB-XL & 0.775$\pm$0.015&0.875$\pm$0.005&0.877$\pm$0.004&0.778$\pm$0.138&0.876$\pm$0.005&0.874$\pm$0.005&0.877$\pm$0.006&\textbf{0.880$\pm$0.005}\\
Ningbo & 0.718$\pm$0.053&0.916$\pm$0.005&0.913$\pm$0.004&0.909$\pm$0.005&0.906$\pm$0.002&0.913$\pm$0.002&0.915$\pm$0.004&\textbf{0.925$\pm$0.001}\\
Chapman & 0.749$\pm$0.052&0.897$\pm$0.003&0.900$\pm$0.003&0.898$\pm$0.008&0.900$\pm$0.004&0.899$\pm$0.002&0.901$\pm$0.007&\textbf{0.912$\pm$0.002}\\
\midrule
\multicolumn{9}{c}{\textbf{Marco $G_{beta}$ score} (The greater, the better)}\\
\midrule
G12EC & 0.339$\pm$0.004&0.452$\pm$0.006&0.460$\pm$0.004&0.448$\pm$0.011&0.450$\pm$0.009&0.447$\pm$0.005&0.465$\pm$0.007&\textbf{0.477$\pm$0.003}\\
PTB-XL & 0.352$\pm$0.008&0.454$\pm$0.007&0.453$\pm$0.010&0.393$\pm$0.080&0.460$\pm$0.003&0.450$\pm$0.009&0.458$\pm$0.003&\textbf{0.467$\pm$0.009}\\
Ningbo & 0.360$\pm$0.027&0.544$\pm$0.011&0.541$\pm$0.007&0.545$\pm$0.010&0.536$\pm$0.004&0.542$\pm$0.006&0.544$\pm$0.006&\textbf{0.563$\pm$0.001}\\
Chapman & 0.368$\pm$0.053&0.523$\pm$0.012&0.526$\pm$0.015&0.523$\pm$0.024&0.521$\pm$0.012&0.523$\pm$0.013&0.530$\pm$0.017&\textbf{0.554$\pm$0.009}\\
\bottomrule
\end{tabular*}
}
\end{center}
\end{table*}

\begin{table*}[h]
\setlength{\tabcolsep}{0.86em}
\begin{center}
\caption{Comparison results between ECGMatch and the state-of-the-art models using the cross-dataset protocol. The mean performance and standard deviations on four databases are shown across three seeds.}
\label{tab:Crosscompare}
\scalebox{1}{
\color{black}
\begin{tabular*}{\linewidth}{@{}lccccccccc@{}}
\toprule
Methods & MixMatch\cite{berthelot2019mixmatch} & FixMatch\cite{sohn2020fixmatch} & FlexMatch\cite{zhang2021flexmatch} & DST\cite{chen2022debiased} & PerMatch\cite{huang2022percentmatch} & SoftMatch\cite{chen2023softmatch} & UPS\cite{rizve2021defense} & \textbf{ECGMatch} \\
\midrule
\multicolumn{9}{c}{\textbf{Ranking loss} (The smaller, the better)}\\
\midrule
G12EC & 0.333$\pm$0.026&0.290$\pm$0.020&0.243$\pm$0.023&0.251$\pm$0.009&0.255$\pm$0.018&0.281$\pm$0.033&0.248$\pm$0.011&\textbf{0.203$\pm$0.004}\\
PTB-XL & 0.474$\pm$0.071&0.285$\pm$0.015&0.254$\pm$0.009&0.265$\pm$0.009&0.272$\pm$0.016&0.271$\pm$0.022&0.270$\pm$0.019&\textbf{0.248$\pm$0.005}\\
Ningbo & 0.319$\pm$0.222&0.148$\pm$0.029&0.138$\pm$0.049&0.127$\pm$0.014&0.156$\pm$0.026&0.169$\pm$0.009&0.160$\pm$0.035&\textbf{0.102$\pm$0.006}\\
Chapman & 0.212$\pm$0.016&0.147$\pm$0.040&0.126$\pm$0.017&0.126$\pm$0.027&0.169$\pm$0.053&0.169$\pm$0.042&0.156$\pm$0.043&\textbf{0.068$\pm$0.002}\\
\midrule
\multicolumn{9}{c}{\textbf{Hamming loss} (The smaller, the better)}\\
\midrule
G12EC & 0.380$\pm$0.003&0.343$\pm$0.007&0.337$\pm$0.006&0.357$\pm$0.007&0.350$\pm$0.009&0.352$\pm$0.015&0.349$\pm$0.006&\textbf{0.331$\pm$0.007}\\
PTB-XL & 0.517$\pm$0.114&0.365$\pm$0.019&0.362$\pm$0.014&0.360$\pm$0.029&0.345$\pm$0.019&0.372$\pm$0.020&0.359$\pm$0.009&\textbf{0.310$\pm$0.001}\\
Ningbo & 0.353$\pm$0.056&0.287$\pm$0.012&0.308$\pm$0.022&0.280$\pm$0.028&0.307$\pm$0.018&0.315$\pm$0.002&0.288$\pm$0.011&\textbf{0.253$\pm$0.008}\\
Chapman & 0.273$\pm$0.013&0.259$\pm$0.008&0.276$\pm$0.011&0.267$\pm$0.014&0.277$\pm$0.014&0.275$\pm$0.003&0.262$\pm$0.015&\textbf{0.219$\pm$0.003}\\
\midrule
\multicolumn{9}{c}{\textbf{Coverage} (The smaller, the better)}\\
\midrule
G12EC & 2.892$\pm$0.094&2.740$\pm$0.078&2.586$\pm$0.092&2.599$\pm$0.020&2.619$\pm$0.071&2.728$\pm$0.125&2.594$\pm$0.036&\textbf{2.415$\pm$0.016}\\
PTB-XL & 3.206$\pm$0.275&2.512$\pm$0.069&2.400$\pm$0.037&2.432$\pm$0.038&2.451$\pm$0.050&2.454$\pm$0.084&2.445$\pm$0.073&\textbf{2.379$\pm$0.023}\\
Ningbo & 2.822$\pm$0.838&2.160$\pm$0.121&2.131$\pm$0.198&2.084$\pm$0.070&2.196$\pm$0.104&2.245$\pm$0.036&2.208$\pm$0.150&\textbf{1.971$\pm$0.025}\\
Chapman & 2.352$\pm$0.067&2.142$\pm$0.176&2.046$\pm$0.068&2.035$\pm$0.095&2.217$\pm$0.206&2.235$\pm$0.173&2.183$\pm$0.181&\textbf{1.803$\pm$0.008}\\
\midrule
\multicolumn{9}{c}{\textbf{MAP} (The greater, the better)}\\
\midrule
G12EC & 0.591$\pm$0.012&0.616$\pm$0.013&0.632$\pm$0.005&0.630$\pm$0.010&0.622$\pm$0.007&0.621$\pm$0.005&0.630$\pm$0.004&\textbf{0.657$\pm$0.009}\\
PTB-XL & 0.518$\pm$0.030&0.532$\pm$0.007&0.551$\pm$0.006&0.538$\pm$0.006&0.558$\pm$0.004&0.545$\pm$0.008&0.553$\pm$0.009&\textbf{0.591$\pm$0.012}\\
Ningbo & 0.560$\pm$0.067&0.663$\pm$0.006&0.665$\pm$0.003&0.667$\pm$0.004&0.649$\pm$0.001&0.658$\pm$0.003&0.667$\pm$0.003&\textbf{0.689$\pm$0.002}\\
Chapman & 0.702$\pm$0.005&0.730$\pm$0.007&0.726$\pm$0.005&0.727$\pm$0.005&0.710$\pm$0.001&0.728$\pm$0.004&0.726$\pm$0.005&\textbf{0.748$\pm$0.004}\\
\midrule
\multicolumn{9}{c}{\textbf{Marco AUC} (The greater, the better)}\\
\midrule
G12EC & 0.755$\pm$0.008&0.779$\pm$0.007&0.789$\pm$0.005&0.784$\pm$0.009&0.781$\pm$0.008&0.783$\pm$0.004&0.787$\pm$0.001&\textbf{0.805$\pm$0.004}\\
PTB-XL & 0.733$\pm$0.035&0.767$\pm$0.008&0.780$\pm$0.004&0.771$\pm$0.002&0.780$\pm$0.006&0.773$\pm$0.008&0.779$\pm$0.010&\textbf{0.800$\pm$0.010}\\
Ningbo & 0.810$\pm$0.045&0.869$\pm$0.003&0.867$\pm$0.001&0.867$\pm$0.002&0.864$\pm$0.003&0.866$\pm$0.003&0.872$\pm$0.001&\textbf{0.874$\pm$0.002}\\
Chapman & 0.864$\pm$0.005&0.888$\pm$0.004&0.889$\pm$0.004&0.889$\pm$0.002&0.880$\pm$0.001&0.888$\pm$0.000&0.890$\pm$0.002&\textbf{0.900$\pm$0.002}\\
\midrule
\multicolumn{9}{c}{\textbf{Marco $G_{beta}$ score} (The greater, the better)}\\
\midrule
G12EC & 0.376$\pm$0.002&0.387$\pm$0.005&0.394$\pm$0.001&0.388$\pm$0.004&0.390$\pm$0.005&0.389$\pm$0.005&0.392$\pm$0.005&\textbf{0.403$\pm$0.002}\\
PTB-XL & 0.312$\pm$0.032&0.337$\pm$0.003&0.346$\pm$0.006&0.345$\pm$0.005&0.353$\pm$0.004&0.347$\pm$0.009&0.347$\pm$0.011&\textbf{0.369$\pm$0.001}\\
Ningbo & 0.380$\pm$0.028&0.419$\pm$0.007&0.413$\pm$0.014&0.429$\pm$0.006&0.408$\pm$0.008&0.408$\pm$0.003&0.426$\pm$0.003&\textbf{0.442$\pm$0.003}\\
Chapman & 0.463$\pm$0.003&0.484$\pm$0.006&0.470$\pm$0.014&0.489$\pm$0.004&0.456$\pm$0.009&0.467$\pm$0.003&0.482$\pm$0.009&\textbf{0.516$\pm$0.006}\\
\bottomrule
\end{tabular*}
}
\end{center}
\end{table*}

\begin{table*}[h]
\setlength{\tabcolsep}{0.73em}
\begin{center}
\caption{Comparison results between ECGMatch and the state-of-the-art models using the mix-dataset protocol. The mean performance and standard deviations on four databases are shown across three seeds.}
\label{tab:mixcompare}
\scalebox{1}{
\color{black}
\begin{tabular*}{\linewidth}{@{}lccccccccc@{}}
\toprule
Methods & MixMatch\cite{berthelot2019mixmatch} & FixMatch\cite{sohn2020fixmatch} & FlexMatch\cite{zhang2021flexmatch} & DST\cite{chen2022debiased} & PerMatch\cite{huang2022percentmatch} & SoftMatch\cite{chen2023softmatch} & UPS\cite{rizve2021defense} & \textbf{ECGMatch} \\
\midrule
Ranking loss & 0.241$\pm$0.057&0.233$\pm$0.026&0.205$\pm$0.014&0.189$\pm$0.023&0.227$\pm$0.031&0.236$\pm$0.033&0.181$\pm$0.012&\textbf{0.150$\pm$0.001}\\
Hamming loss & 0.313$\pm$0.014&0.292$\pm$0.006&0.299$\pm$0.009&0.295$\pm$0.007&0.307$\pm$0.014&0.303$\pm$0.006&0.288$\pm$0.004&\textbf{0.270$\pm$0.001}\\
Coverage & 2.462$\pm$0.218&2.437$\pm$0.102&2.322$\pm$0.061&2.265$\pm$0.098&2.411$\pm$0.125&2.445$\pm$0.114&2.230$\pm$0.057&\textbf{2.101$\pm$0.009}\\
MAP & 0.625$\pm$0.021&0.643$\pm$0.009&0.643$\pm$0.009&0.645$\pm$0.011&0.635$\pm$0.013&0.647$\pm$0.005&0.640$\pm$0.009&\textbf{0.658$\pm$0.006}\\
Marco AUC & 0.827$\pm$0.011&0.834$\pm$0.005&0.832$\pm$0.004&0.836$\pm$0.006&0.831$\pm$0.004&0.835$\pm$0.003&0.834$\pm$0.005&\textbf{0.838$\pm$0.003}\\
Marco $G_{beta}$ & 0.417$\pm$0.008&0.431$\pm$0.003&0.432$\pm$0.004&0.435$\pm$0.003&0.428$\pm$0.008&0.431$\pm$0.007&0.434$\pm$0.004&\textbf{0.442$\pm$0.002}\\
\bottomrule
\end{tabular*}
}
\end{center}
\end{table*}

\subsection{Ablation Study}\label{ablation}
In order to quantitatively assess the contribution of different modules in the ECGMatch, we successively remove one of them and evaluate the model performance using the three established protocols. Table \ref{tab:ablation_within}, Table \ref{tab:ablation_cross}, Table \ref{tab:ablation_mix} report the ablation studies on different experimental protocols.  1) When the pseudo-label generation module is removed ($\lambda_u=0$, Eq.\ref{Eq:finalloss}), the performance of the proposed model decreases in all the experimental protocols, which demonstrates the advantages of introducing pseudo-labels for semi-supervised learning. For example, in the within-dataset protocols (Table \ref{tab:ablation_within}), the hamming loss on the Chapman dataset increases from 0.139$\pm$0.002 to 0.163$\pm$0.009 while the MAP decreases from 0.775$\pm$0.014 to 0.761$\pm$0.010. Notably, as the parameter $\lambda_u$ is set to zero, the following refinement module is also disabled. 2) The significant negative effect of removing the pseudo-label refinement module is observed in the results. In the cross-dataset protocol (Table \ref{tab:ablation_cross}), the hamming loss on the Chapman dataset increases from 0.219$\pm$0.003 to 0.242$\pm$0.007 while the MAP drops from 0.748$\pm$0.004 to 0.732$\pm$0.006. This phenomenon indicates that increasing the importance of the trust-worthy pseudo-labels in loss computation greatly enhances the model performance. 3) Comparing the results with and without the label correlation alignment module ($\lambda_f=0$, Eq.\ref{Eq:finalloss}), a significant performance drop is observed when the module is removed. In the mix-dataset protocols (Table \ref{tab:ablation_mix}), the hamming loss increases from 0.270$\pm$0.001 to 0.282$\pm$0.010 while the MAP decreases from 0.658$\pm$0.006 to 0.640$\pm$0.010. This phenomenon demonstrates the benefits of capturing the correlation between different categories, which has also been reported in the other multi-label learning studies\cite{zhang2007ml,liu2017easy,wang2021semi}.

\begin{table*}[h]
\setlength{\tabcolsep}{0.8em}
\begin{center}
\caption{The ablation study of the proposed ECGMatch (with-dataset protocol).}
\label{tab:ablation_within}
\scalebox{1}{
\color{black}
\begin{tabular*}{\hsize}{@{}@{\extracolsep{\fill}}lcccc@{}}
\toprule
Methods   & G12EC  & PTB & Ningbo & Chapman\\
\midrule
\multicolumn{5}{c}{\textbf{Ranking loss} (The smaller, the better)}\\
\midrule
Without pseudo-label generation & 0.145$\pm$0.008&0.140$\pm$0.004&0.053$\pm$0.004&0.060$\pm$0.003\\
Without pseudo-label refinement & 0.164$\pm$0.012&0.140$\pm$0.005&0.046$\pm$0.002&0.066$\pm$0.013\\
Without label correlation alignment  & 0.150$\pm$0.010&0.138$\pm$0.007&0.063$\pm$0.010&0.061$\pm$0.002\\
\textbf{ECGMatch} & \textbf{0.140$\pm$0.006}&\textbf{0.134$\pm$0.003}&\textbf{0.045$\pm$0.002}&\textbf{0.052$\pm$0.002}\\
\midrule
\multicolumn{5}{c}{\textbf{Hamming loss} (The smaller, the better)}\\
\midrule
Without pseudo-label generation & 0.300$\pm$0.014&0.244$\pm$0.006&0.127$\pm$0.006&0.163$\pm$0.009\\
Without pseudo-label refinement & 0.296$\pm$0.009&0.241$\pm$0.013&0.138$\pm$0.012&0.163$\pm$0.014\\
Without label correlation alignment  & 0.290$\pm$0.007&0.243$\pm$0.005&0.131$\pm$0.003&0.151$\pm$0.011\\
\textbf{ECGMatch} & \textbf{0.278$\pm$0.008}&\textbf{0.233$\pm$0.009}&\textbf{0.122$\pm$0.001}&\textbf{0.139$\pm$0.002}\\
\midrule
\multicolumn{5}{c}{\textbf{Coverage} (The smaller, the better)}\\
\midrule
Without pseudo-label generation & 2.192$\pm$0.031&1.946$\pm$0.019&1.758$\pm$0.018&1.795$\pm$0.010\\
Without pseudo-label refinement & 2.229$\pm$0.047&1.946$\pm$0.021&1.722$\pm$0.011&1.822$\pm$0.055\\
Without label correlation alignment  & 2.220$\pm$0.048&1.939$\pm$0.032&1.804$\pm$0.046&1.803$\pm$0.013\\
\textbf{ECGMatch} & \textbf{2.173$\pm$0.027}&\textbf{1.922$\pm$0.015}&\textbf{1.724$\pm$0.010}&\textbf{1.761$\pm$0.021}\\
\midrule
\multicolumn{5}{c}{\textbf{MAP} (The greater, the better)}\\
\midrule
Without pseudo-label generation & 0.728$\pm$0.006&0.739$\pm$0.009&0.801$\pm$0.002&0.761$\pm$0.010\\
Without pseudo-label refinement & 0.725$\pm$0.018&0.741$\pm$0.012&0.805$\pm$0.004&0.739$\pm$0.040\\
Without label correlation alignment  & 0.731$\pm$0.007&0.741$\pm$0.010&0.794$\pm$0.006&0.751$\pm$0.006\\
\textbf{ECGMatch} & \textbf{0.742$\pm$0.005}&\textbf{0.748$\pm$0.009}&\textbf{0.808$\pm$0.001}&\textbf{0.775$\pm$0.014}\\
\midrule
\multicolumn{5}{c}{\textbf{Marco AUC} (The greater, the better)}\\
\midrule
Without pseudo-label generation & 0.849$\pm$0.005&0.878$\pm$0.004&0.920$\pm$0.001&0.905$\pm$0.004\\
Without pseudo-label refinement & 0.852$\pm$0.003&0.877$\pm$0.007&0.922$\pm$0.002&0.906$\pm$0.005\\
Without label correlation alignment  & 0.851$\pm$0.005&0.879$\pm$0.005&0.915$\pm$0.004&0.906$\pm$0.003\\
\textbf{ECGMatch} & \textbf{0.854$\pm$0.003}&\textbf{0.880$\pm$0.005}&\textbf{0.925$\pm$0.001}&\textbf{0.912$\pm$0.002}\\
\midrule
\multicolumn{5}{c}{\textbf{Marco $G_{beta}$ score} (The greater, the better)}\\
\midrule
Without pseudo-label generation & 0.467$\pm$0.010&0.463$\pm$0.007&0.553$\pm$0.005&0.538$\pm$0.020\\
Without pseudo-label refinement & 0.460$\pm$0.006&0.465$\pm$0.012&0.539$\pm$0.016&0.539$\pm$0.011\\
Without label correlation alignment  & 0.467$\pm$0.011&0.463$\pm$0.005&0.544$\pm$0.002&0.541$\pm$0.019\\
\textbf{ECGMatch} & \textbf{0.477$\pm$0.003}&\textbf{0.467$\pm$0.009}&\textbf{0.563$\pm$0.001}&\textbf{0.554$\pm$0.009}\\
\bottomrule
\end{tabular*}
}
\end{center}
\end{table*}

\begin{table*}[h]
\setlength{\tabcolsep}{0.8em}
\begin{center}
\caption{The ablation study of the proposed ECGMatch (cross-dataset protocol).}
\label{tab:ablation_cross}
\scalebox{1}{
\color{black}
\begin{tabular*}{\hsize}{@{}@{\extracolsep{\fill}}lcccc@{}}
\toprule
Methods   & G12EC  & PTB & Ningbo & Chapman\\
\midrule
\multicolumn{5}{c}{\textbf{Ranking loss} (The smaller, the better)}\\
\midrule
Without pseudo-label generation & 0.215$\pm$0.012&0.259$\pm$0.007&0.109$\pm$0.005&0.082$\pm$0.007\\
Without pseudo-label refinement & 0.225$\pm$0.011&0.273$\pm$0.010&0.143$\pm$0.015&0.079$\pm$0.000\\
Without label correlation alignment  & 0.226$\pm$0.019&0.252$\pm$0.002&0.126$\pm$0.010&0.101$\pm$0.009\\
\textbf{ECGMatch} & \textbf{0.203$\pm$0.004}&\textbf{0.248$\pm$0.005}&\textbf{0.102$\pm$0.006}&\textbf{0.068$\pm$0.002}\\
\midrule
\multicolumn{5}{c}{\textbf{Hamming loss} (The smaller, the better)}\\
\midrule
Without pseudo-label generation & 0.338$\pm$0.002&0.355$\pm$0.010&0.271$\pm$0.006&0.242$\pm$0.007\\
Without pseudo-label refinement & 0.358$\pm$0.008&0.350$\pm$0.014&0.302$\pm$0.009&0.266$\pm$0.030\\
Without label correlation alignment  & 0.344$\pm$0.001&0.341$\pm$0.014&0.281$\pm$0.010&0.270$\pm$0.003\\
\textbf{ECGMatch} & \textbf{0.331$\pm$0.007}&\textbf{0.310$\pm$0.001}&\textbf{0.253$\pm$0.008}&\textbf{0.219$\pm$0.003}\\
\midrule
\multicolumn{5}{c}{\textbf{Coverage} (The smaller, the better)}\\
\midrule
Without pseudo-label generation & 2.455$\pm$0.038&2.425$\pm$0.026&1.994$\pm$0.014&1.867$\pm$0.032\\
Without pseudo-label refinement & 2.515$\pm$0.020&2.510$\pm$0.080&2.144$\pm$0.129&1.816$\pm$0.012\\
Without label correlation alignment  & 2.498$\pm$0.067&2.393$\pm$0.014&2.085$\pm$0.037&1.949$\pm$0.033\\
\textbf{ECGMatch} & \textbf{2.415$\pm$0.016}&\textbf{2.379$\pm$0.023}&\textbf{1.971$\pm$0.025}&\textbf{1.803$\pm$0.008}\\
\midrule
\multicolumn{5}{c}{\textbf{MAP} (The greater, the better)}\\
\midrule
Without pseudo-label generation & 0.643$\pm$0.006&0.568$\pm$0.006&0.665$\pm$0.007&0.732$\pm$0.006\\
Without pseudo-label refinement & 0.635$\pm$0.013&0.571$\pm$0.013&0.664$\pm$0.018&0.739$\pm$0.008\\
Without label correlation alignment  & 0.639$\pm$0.010&0.572$\pm$0.011&0.670$\pm$0.003&0.718$\pm$0.006\\
\textbf{ECGMatch} & \textbf{0.657$\pm$0.009}&\textbf{0.591$\pm$0.012}&\textbf{0.689$\pm$0.002}&\textbf{0.748$\pm$0.004}\\
\midrule
\multicolumn{5}{c}{\textbf{Marco AUC} (The greater, the better)}\\
\midrule
Without pseudo-label generation & 0.796$\pm$0.002&0.782$\pm$0.007&0.866$\pm$0.002&0.890$\pm$0.005\\
Without pseudo-label refinement & 0.788$\pm$0.007&0.790$\pm$0.003&0.865$\pm$0.004&0.889$\pm$0.005\\
Without label correlation alignment  & 0.792$\pm$0.004&0.794$\pm$0.011&0.869$\pm$0.001&0.884$\pm$0.004\\
\textbf{ECGMatch} & \textbf{0.805$\pm$0.004}&\textbf{0.800$\pm$0.010}&\textbf{0.874$\pm$0.002}&\textbf{0.900$\pm$0.002}\\
\midrule
\multicolumn{5}{c}{\textbf{Marco $G_{beta}$ score} (The greater, the better)}\\
\midrule
Without pseudo-label generation & 0.397$\pm$0.003&0.357$\pm$0.007&0.441$\pm$0.005&0.497$\pm$0.005\\
Without pseudo-label refinement & 0.390$\pm$0.003&0.355$\pm$0.001&0.421$\pm$0.001&0.506$\pm$0.017\\
Without label correlation alignment  & 0.392$\pm$0.002&0.357$\pm$0.011&0.430$\pm$0.005&0.476$\pm$0.008\\
\textbf{ECGMatch} & \textbf{0.403$\pm$0.002}&\textbf{0.369$\pm$0.001}&\textbf{0.442$\pm$0.003}&\textbf{0.516$\pm$0.006}\\
\bottomrule
\end{tabular*}
}
\end{center}
\end{table*}

\begin{table*}[h]
\begin{center}
\caption{The ablation study of the proposed ECGMatch (mix-dataset protocols).}
\label{tab:ablation_mix}
\scalebox{1}{
\color{black}
\begin{tabular*}{\hsize}{@{}@{\extracolsep{\fill}}lcccccc@{}}
\toprule
Methods & Ranking loss & Hamming loss & Coverage & MAP & Marco AUC & Marco $G_{beta}$ score \\
\midrule
Without pseudo-label generation & 0.156$\pm$0.003&0.286$\pm$0.005&2.122$\pm$0.015&0.651$\pm$0.007&0.837$\pm$0.003&0.438$\pm$0.002\\
Without pseudo-label refinement & 0.165$\pm$0.011&0.285$\pm$0.006&2.163$\pm$0.048&0.651$\pm$0.004&0.829$\pm$0.005&0.436$\pm$0.004\\
Without label correlation alignment & 0.169$\pm$0.004&0.282$\pm$0.010&2.183$\pm$0.019&0.640$\pm$0.010&0.832$\pm$0.004&0.432$\pm$0.005\\
\textbf{ECGMatch} & \textbf{0.150$\pm$0.001}&\textbf{0.270$\pm$0.001}&\textbf{2.101$\pm$0.009}&\textbf{0.658$\pm$0.006}&\textbf{0.838$\pm$0.003}&\textbf{0.442$\pm$0.002}\\
\bottomrule
\end{tabular*}
}
\end{center}
\end{table*}

\subsection{Comparison of Different Augmentation Strategies}
\begin{figure*}[h]
  \begin{center}
  \subfloat[Within-dataset protocol]{\includegraphics[width=1\textwidth]{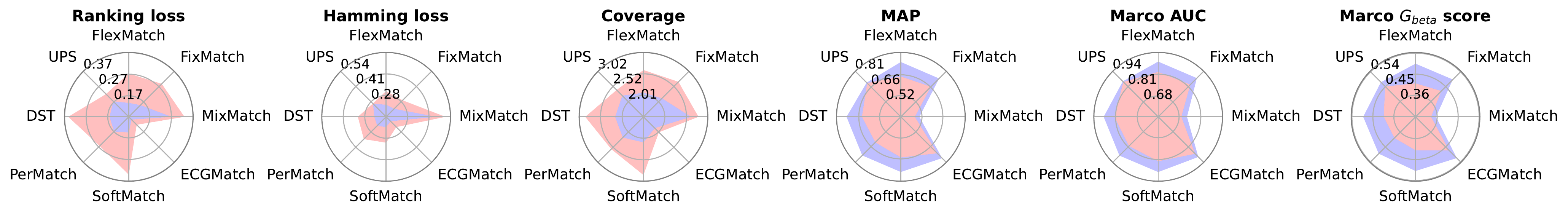}}
    \\
  \subfloat[Cross-dataset protocol]{\includegraphics[width=1\textwidth]{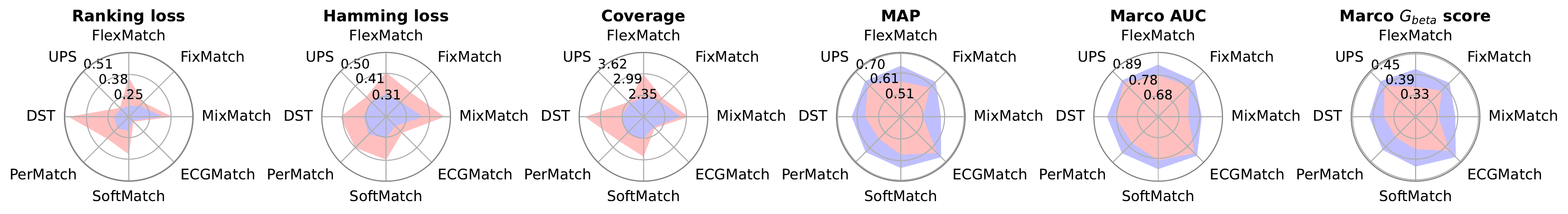}}
  \\
  \subfloat[Mix-dataset protocol]{\includegraphics[width=1\textwidth]{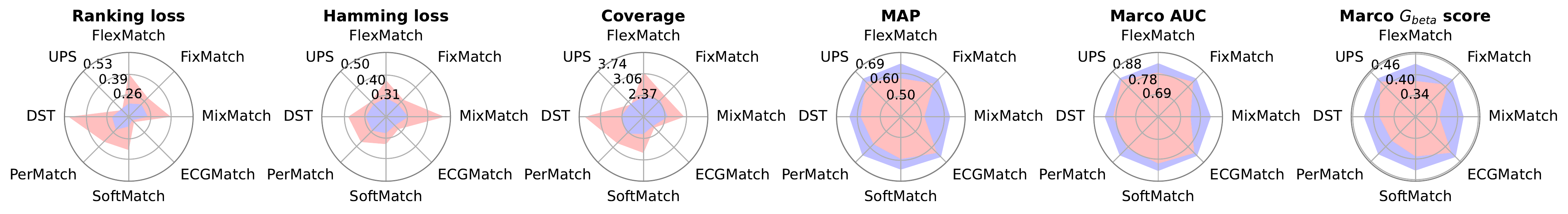}}
  \caption{Performance comparison of two augmentation pipelines using radar charts. The vertices of the red zone denote the performance of the model with the fixed sequential perturbations, while the vertices of the blue zone represent the performance of the model with the proposed ECGAugment.}
  \label{fig:augment}
  \end{center}
\end{figure*}

In this section, we further investigate the effectiveness of the ECGAugment module in ECG signal augmentation.  Using the aforementioned protocols, we compare its impact on model performance with the fixed sequential perturbations used in previous studies\cite{kiyasseh2021clocs,nonaka2020electrocardiogram}. The averaged performance across four datasets is shown in Fig. \ref{fig:augment}, where an obvious performance enhancement attributed to the ECGAugment could be observed from the performance of different models. For the evaluation metrics where smaller is better, the blue zones (ECGAugment) in the radar charts are surrounded by the red zones (fixed sequential perturbations). Conversely,  for the metrics where greater is better, the blue zones cover the red zones. These phenomena demonstrate the superiority of the proposed ECGAugment in the downstream classification tasks. In other words, it increases the sample diversity by enhancing the randomness in data augmentation, which can improve the model performance\cite{UDA2020,cubuk2020randaugment,gontijo2020affinity}.

\subsection{Statistical Analysis}
\begin{figure*}[h]
  \begin{center}
  \subfloat[Within-dataset protocol]{\includegraphics[width=1\textwidth]{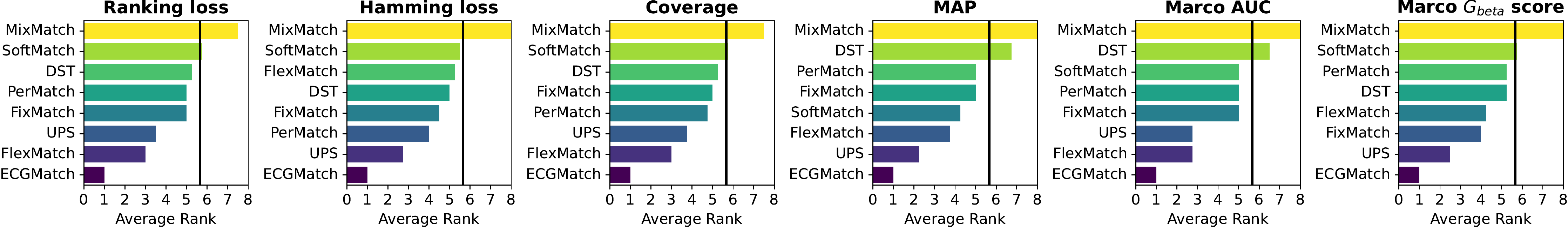}}
    \\
  \subfloat[Cross-dataset protocol]{\includegraphics[width=1\textwidth]{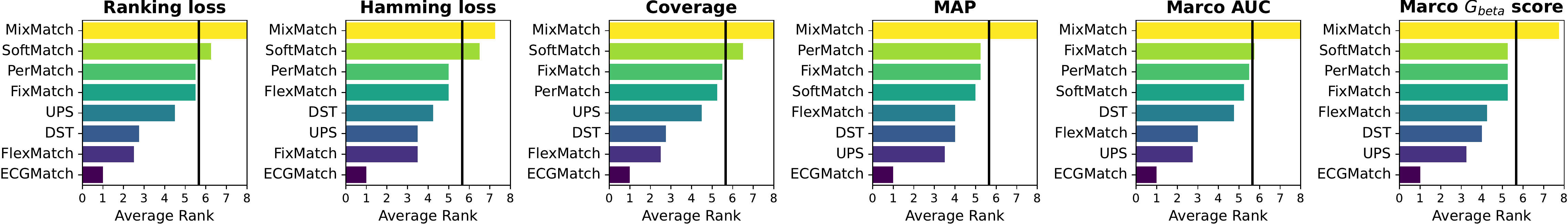}}
  \caption{Comparison of ECGMatch against other compared models based on the Bonferroni-Dunn test (cross-dataset protocol).  ECGMatch is deemed to have a significantly better performance than one compared model if their
average ranks differ by at least one \emph{critical difference}=4.6592, as denoted by the intersection of the bar with the black lines.}
  \label{fig:Dunn_test}
  \end{center}
\end{figure*}

To statistically analyze the performance difference between the ECGMatch and other SOTA models, a commonly used \emph{Friedman test} and the post-hoc \emph{Bonferroni-Dunn test} are employed. Following the pipeline of the aforementioned tests\cite{demvsar2006statistical}, we use the performances of different models in the within-dataset protocol and cross-dataset protocol for comparison. Table \ref{tab:Friedman} presents the Friedman statistics $F_F$ and the associated critical value for each metric (comparing models $k=8$, datasets $N=4$). Based on the results ($F_F$\textgreater 3.2590), we can reject the null hypothesis that the compared models show no significant difference in performance at a 0.05 significance level. Then the post-hoc \emph{Bonferroni-Dunn test} are applied to describe the performance gap between the control model (ECGMatch) and the other models. For each evaluation metric, we calculate the average rank of all the models across four datasets and determine the rank differences between the control model and the other compared models. Note that the top-performing model is assigned a rank of 1, and the second-best model gets a rank of 2, and so on. The control model (ECGMatch) is significantly better than one compared model if their rank difference is larger than at least one \emph{critical difference} (CD=4.6592 in our experiment). Fig.\ref{fig:Dunn_test} presents the mean rank of different models on different evaluation metrics. It is evident that the proposed ECGMatch ranks the best in terms of all the metrics and outperforms some competitors at a 0.05 significance level, such as MixMatch\cite{berthelot2019mixmatch}, DST\cite{chen2022debiased} and SoftMatch\cite{chen2023softmatch}. In summary, these statistical results convincingly demonstrate the superiority of the proposed ECGMatch.

\begin{table*}[h]
\caption{\emph{Friedman statistics} $F_F$ for each metric and the critical value at 0.05 significance level (the number of comparing models $k=8$ and datasets $N=4$).}
\label{tab:Friedman}
\color{black}
\begin{tabular*}{\textwidth}{@{}@{\extracolsep{\fill}}lcclcclcc@{}}
\toprule
\multicolumn{9}{c}{\textbf{Within-dataset protocol}}\\
\midrule
Evaluation metric & $F_F$ & critical values & Evaluation metric & $F_F$ & critical values & Evaluation metric & $F_F$ & critical values\\
\midrule
Ranking loss & 05.4706 &3.2590 &MAP & 17.1600 &3.2590&Coverage & 05.1290 &3.2590\\
Hamming loss & 07.1818 &3.2590 &Marco AUC & 16.0189 &3.2590&Marco $G_{beta}$ & 09.0000 &3.2590\\
\midrule
\multicolumn{9}{c}{\textbf{Cross-dataset protocol}}\\
\midrule
Evaluation metric & $F_F$ & critical values & Evaluation metric & $F_F$ & critical values & Evaluation metric & $F_F$ & critical values\\
\midrule
Ranking loss & 20.4419 &3.2590 &MAP & 05.6154 &3.2590&Coverage & 22.8462 &3.2590\\
Hamming loss & 05.0640 &3.2590 &Marco AUC & 11.0000 &3.2590&Marco $G_{beta}$ & 05.0640 &3.2590\\
\bottomrule
\end{tabular*}
\end{table*}

\subsection{Sensitivity Analysis}
\begin{figure*}[h]
  \begin{center}
  \subfloat[Within-dataset protocol]{\includegraphics[width=1\textwidth]{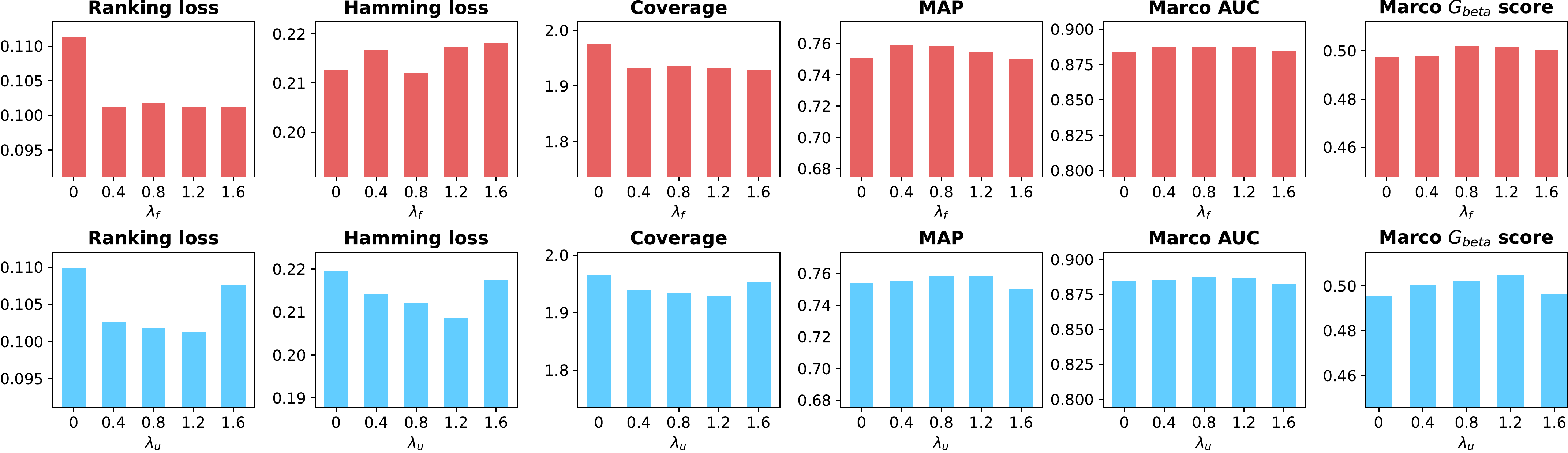}}
    \\
  \subfloat[Cross-dataset protocol]{\includegraphics[width=1\textwidth]{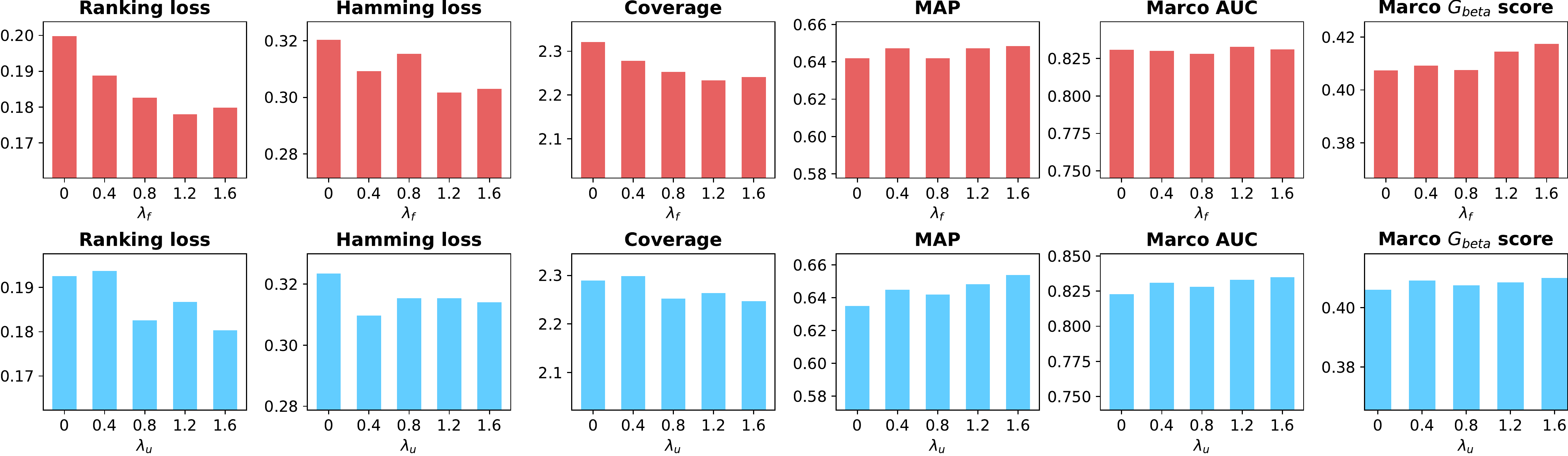}}
  \\
  \subfloat[Mix-dataset protocol]{\includegraphics[width=1\textwidth]{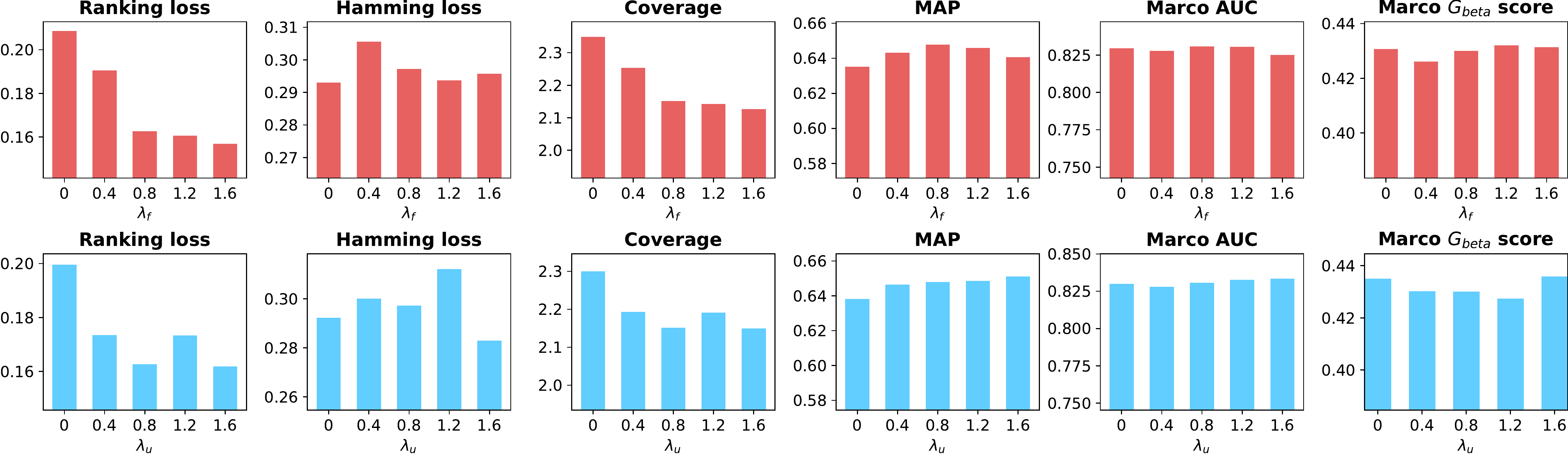}}
  \caption{Average model performance in different protocols under varying hyperparameters}
  \label{fig:hyperparameter}
  \end{center}
\end{figure*}

In this section, we use a grid-search method to investigate the impact of varying hyper-parameters on the performance of the proposed model. For simplicity, we only focus on two critical hyper-parameters $\lambda_u$ and $\lambda_f$ in Eq. \ref{Eq:finalloss}. Specifically, $\lambda_u$ controls the weight of the unsupervised binary cross entropy loss $L_u$, while $\lambda_f$ controls the weight of the label correlation alignment loss $L_f$. In the grid search process, we adjust the values of the hyper-parameters and use different evaluation protocols to evaluate the average model performance across four datasets. First, we fix $\lambda_u$ at 0.8 and adjust $\lambda_f$ from 0 to 1.6 in steps of 0.4. Then, we fix $\lambda_f$ at 0.8 and adjust $\lambda_u$ in the same manner. As illustrated in Fig. \ref{fig:hyperparameter}, the performance of the proposed ECGMatch in each evaluation metric is relatively insensitive to the changes of the two hyper-parameters, which suggests its stability in clinical applications.

\section{Conclusion}
In this study, we point out three important real-world challenges in ECG-based CVDs prediction: 1) Label scarcity problem. 2) Poor performance on unseen datasets. 3) Co-occurrence of multiple CVDs. To address the challenges simultaneously, we propose a novel framework (ECGMatch) that combines data augmentation, pseudo-label learning, and label correlation alignment modules to formulate a unified framework. Further, we re-annotate four public datasets and propose three practical experimental protocols to conduct a multi-dataset evaluation of the proposed model. Extensive experiments on three protocols and four datasets convincingly demonstrated the superiority of the proposed model against other SOTA models. We believe the ECGMatch can provide a reliable baseline for future research on ECG-based CVDs prediction. However, the class imbalance problem in the ECG datasets continues to pose a significant challenge. Therefore, we advocate for future research on this ongoing issue. 


%
\bibliographystyle{IEEEtran}
\bibliography{references}
\ifCLASSOPTIONcompsoc
  \section*{Acknowledgments}
\else
  \section*{Acknowledgment}
\fi

This work was supported by an NIHR Research Professorship, an RAEng Research Chair, the InnoHK Hong Kong projects under the Hong Kong Center for Cerebro-cardiovascular Health Engineering (COCHE), the NIHR Oxford Biomedical Research Centre (BRC), and the Pandemic Sciences Institute at the University of Oxford.
\appendices
\section{Re-annotation of the public datasets}\label{sec:dataset}
As different public datasets have different kinds of abnormal ECG signals and adopt various standards for CVDs diagnosis, there is a strong label distribution shift among them. To enable a multi-dataset evaluation of different models, we re-annotate the ECG recordings from the datasets using the same annotation scheme. As shown in Table \ref{tab:anotation}, a comparison between the original and our annotations is presented. Specifically, we download the original annotations of the recordings from the Physionet\cite{alday2020classification} and assign five new labels to them (Abnormal Rhythms, ST/T Abnormalities, Conduction Disturbance, Other Abnormalities, Normal Signals). Note that each recording can belong to two or more categories simultaneously. The definition of the first three abnormalities originates from the ECG statements of the PTB-XL database\cite{wagner2020ptb}. For the other CVDs which are rare in the four public datasets and difficult to make a general definition, we categorize them as Other Abnormalities. If there are no potential CVDs from a given ECG segment, we regard it as Normal Signals. In practice, the 'Normal' class should not co-occur with other categories and is also considered for label correlation alignment to help our model avoid confusing predictions in which CVDs are detected in normal signals.

It is important to acknowledge that our annotation scheme might not be optimal as diagnosing some CVDs can be complex and uniform definitions are challenging to establish. For example, as the ‘prolonged pr interval’ is one of the criteria to diagnose the ‘1st degree av block’, it is difficult to separate them into two categories\cite{oldroyd2022first}.  However, one might argue that their association is unclear if the pr interval is prolonged but does not cross the 1st av block threshold\cite{pranata2019association}. Consequently, we simply annotate the segments with ‘1st degree av block’ and ‘prolonged pr interval’ as  Conduction Disturbance and Other Abnormalities simultaneously. The segments only with ‘prolonged pr interval’ are only labeled with Other Abnormalities. On the other hand, some labels provided by the Physionet are inaccurate\cite{alday2020classification}, making it difficult to design an optimal re-annotated strategy. After re-annotation, the class distributions of the four databases are shown in Table \ref{tab:class}. It can be observed that there is an obvious class distribution mismatch across different datasets, which challenges the robustness of the SSL models on unseen datasets\cite{oliver2018realistic}. 
\begin{table*}[h]
\setlength{\tabcolsep}{0.75em}
\fontsize{8}{9.5}\selectfont
\caption{A comparison between the original and our annotation  }
\label{tab:anotation}
\color{black}
\begin{tabular*}{\hsize}{@{}lclc@{}}
\toprule
\textbf{Original annotation} & \textbf{Our annotation} &
\textbf{Original annotation} & \textbf{Our annotation}\\
\midrule
atrial fibrillation & Abnormal Rhythms & left bundle branch block
 & Conduction Disturbance\\
atrial flutter  & Abnormal Rhythms & non-specific intraventricular
conduction disorder & Conduction Disturbance\\
bradycardia  & Abnormal Rhythms & right bundle branch block
 & Conduction Disturbance\\
pacing rhythm  & Abnormal Rhythms & av block & Conduction
Disturbance\\
sinus arrhythmia & Abnormal Rhythms & complete heart block &
Conduction Disturbance\\
sinus bradycardia  & Abnormal Rhythms & 2nd degree av block 
& Conduction Disturbance\\
sinus tachycardia & Abnormal Rhythms & mobitz type II
atrioventricular block  & Conduction Disturbance\\
prolonged qt interval & ST/T Abnormalities & incomplete left bundle branch
block & Conduction Disturbance\\
t wave abnormal & ST/T Abnormalities & left posterior fascicular block
 & Conduction Disturbance\\
t wave inversion & ST/T Abnormalities & sinoatrial block &
Conduction Disturbance\\
inferior ischaemia & ST/T Abnormalities & wolff parkinson white pattern & Conduction Disturbance\\
lateral ischaemia & ST/T Abnormalities & left axis deviation & Other
Abnormalities\\
nonspecific st abnormality & ST/T Abnormalities & low qrs voltages
& Other Abnormalities\\
st changes & ST/T Abnormalities & premature atrial contraction &
Other Abnormalities\\
st depression & ST/T Abnormalities & poor R wave progression &
Other Abnormalities\\
st elevation& ST/T Abnormalities & premature ventricular contractions & Other Abnormalities\\
st interval abnormal & ST/T Abnormalities & qwave abnormal & Other
Abnormalities\\
bundle branch block & Conduction Disturbance & right axis
deviation & Other Abnormalities\\
complete left bundle branch block & Conduction Disturbance &
supraventricular premature beats & Other Abnormalities\\
complete right bundle branch block & Conduction Disturbance &
ventricular premature beats & Other Abnormalities\\
1st degree av block & Conduction Disturbance & ventricular
ectopics & Other Abnormalities\\
incomplete right bundle branch block & Conduction Disturbance &
prolonged pr interval & Other Abnormalities\\
left anterior fascicular block & Conduction Disturbance & sinus
rhythm & Normal Signals\\
\bottomrule
\end{tabular*}
\end{table*}

\begin{table*}[h]
\fontsize{8}{9.5}\selectfont
\setlength{\tabcolsep}{0.4em}
\caption{Class distribution of different public datasets after re-annotation.}
\label{tab:class}
\color{black}
\begin{tabular*}{\hsize}{@{}@{\extracolsep{\fill}}lccccc@{}}
\toprule
Datasets & Conduction Disturbance & Abnormal Rhythms & ST/T Abnormalities  & Other Abnormalities & Normal
Signals \\
\midrule
G12EC Database\cite{alday2020classification} &2236&3977&4991&2627&1752\\
PTB-XL database\cite{wagner2020ptb} &4907&4087&4299&7296&6432\\
Chapman databases\cite{zheng202012} &1198&7682&2951&1445&1366\\
Ningbo databases\cite{zheng2020optimal} &3843&28217&10407&5339&4542\\
\bottomrule
\end{tabular*}
\end{table*}

\section{Effect of different similarity metrics}\label{sec:sim}
Recall that the label correlation $\hat{r}_{c_1,c_2}\in[0,1]$ between class $c_1$ and class $c_2$ can be estimated by the cosine similarity (Eq.\ref{Eq:cosine_similarity}) between the label sequences ($y_{c_1}$,$y_{c_2}$) on the two classes. In this section, the effect of different similarity metrics on model performance is also examined. Specifically, we use other metrics to compute the label correlation  $\hat{r}_{c_1,c_2}$, such as the Pearson coefficient (Eq.\ref{Eq:pearson_similarity}) and the Euclidean distance (Eq.\ref{Eq:euclid_similarity}), given as,
\begin{equation}
\label{Eq:pearson_similarity}
\hat{r}_{c_1,c_2}=\hat{\rho}^2,\hat{\rho}=\frac{Cov(y_{c_1},y_{c_2})}{\sigma_{y_{c_1}} \sigma_{y_{c_2}}},
\end{equation}
\begin{equation}
\label{Eq:euclid_similarity}
\hat{r}_{c_1,c_2}=\frac{1}{1+d},d=\left\|y_{c_1}-y_{c_2}\right\|,
\end{equation}
where $\mu_{c_1}$ and $\mu_{c_2}$ are the mean values of the elements in vector $y_{c_1}$ and $y_{c_2}$. Another approach to measure the label correlation is by computing the co-occurrence frequency of two diseases\cite{ge2021multi}. However, we argue that it is not applicable for unlabeled samples lacking binary ground-truth. While it is possible to binarize the generated pseudo-labels using predetermined thresholds, this will introduce additional costs for threshold selection.

As shown in Table \ref{tab:sim_within}, Table \ref{tab:sim_cross} and Table \ref{tab:sim_mix}, the model performance under different similarity metrics is presented. Experiment results on three protocols indicate that cosine similarity outperforms the other metrics, providing an empirical validation of its superiority. Moreover, we provide a theoretical analysis to support the conclusion further. The co-occurrence between two CVDs classes $c_1$ and $c_2$ can be represented by conditional probabilities $P(c_1=1|c_2=1)$ and $P(c_2=1|c_1=1)$, where "$c_1=1$" indicates the existence of CVDs $c_1$. Then we can drive the connection between the cosine similarity and the conditional probabilities, as
\begin{equation}
\label{Eq:cosine_similarity_connect}
\begin{split}  
\hat{r}_{c_1,c_2}&=\frac{y_{c_1}^Ty_{c_2}}{\left\|y_{c_1}\right\|\left\|y_{c_2}\right\|}\approx\frac{NP(c_1=1,c_2=1)}{\sqrt{NP(c_1=1)}\sqrt{NP(c_2=1)}}\\&=\frac{\sqrt{P(c_1=1,c_2=1)}\sqrt{P(c_1=1,c_2=1)}}{\sqrt{P(c_1=1)}\sqrt{P(c_2=1)}}\\&=\sqrt{\frac{P(c_1=1,c_2=1)P(c_1=1,c_2=1)}{P(c_1=1)P(c_2=1)}}\\&=\sqrt{P(c_1=1|c_2=1)P(c_2=1|c_1=1)},
\end{split}
\end{equation}
where $N$ is the number of samples in the label sequences $y_{c_1}$ and $y_{c_2}$. Eq.\ref{Eq:cosine_similarity_connect} shows that the cosine similarity can eliminate the marginal probabilities $P(c_1=1)$ and $P(c_2=1)$, which represent the class distributions of different classes and should not be considered in the computation of $\hat{r}_{c_1,c_2}$. However, other metrics fail to eliminate them, resulting in additional errors and degraded model performance when the class distributions vary across different databases. Specifically, as demonstrated in Eq.\ref{Eq:euclid_similarity_connect} and Eq.\ref{Eq:pearson_similarity_connect}, the label correlation $\hat{r}_{c_1,c_2}$ computed based on the Euclidean distance or the Pearson coefficient is influenced by the class distributions ($P(c_1=1)$ and $P(c_2=1)$), which vary across different databases (Table \ref{tab:class}). This limitation explains their poor performance on the Ningbo and Chapman databases, which exhibit a higher degree of class imbalance than other databases. In summary, experimental and theoretical results show that the ECGMatch is compatible with different similarity metrics and performs better using cosine similarity.
\begin{equation}
\label{Eq:pearson_similarity_connect}
\begin{split} 
&\hat{r}_{c_1,c_2}=\hat{\rho}^2,\hat{\rho}=\frac{(y_{c_1}-\mu_{c_1})^T(y_{c_2}-\mu_{c_2})}{\left\|y_{c_1}-\mu_{c_1}\right\|\left\|y_{c_2}-\mu_{c_2}\right\|}\\&\approx\frac{(y_{c_1}-\mathbf{1}P(c_1=1))^T(y_{c_2}-\mathbf{1}P(c_2=1))}{\left\|y_{c_1}-\mathbf{1}P(c_1=1)\right\|\left\|y_{c_2}-\mathbf{1}P(c_2=1)\right\|}\\&=
\frac{y_{c_1}^Ty_{c_2}-y_{c_1}^T\mathbf{1}P(c_2=1)-\mathbf{1}^TP(c_1=1)y_{c_2}}{\left\|y_{c_1}-\mathbf{1}P(c_1=1)\right\|\left\|y_{c_2}-\mathbf{1}P(c_2=1)\right\|}\\&\hspace{1em}+\frac{NP(c_1=1)P(c_2=1)}{\left\|y_{c_1}-\mathbf{1}P(c_1=1)\right\|\left\|y_{c_2}-\mathbf{1}P(c_2=1)\right\|}\\&=
\frac{P(c_1=1,c_2=1)-P(c_1=1)P(c_2=1)}{\sqrt{P(c_1=1)-P(c_1=1)^2}\sqrt{P(c_2=1)-P(c_2=1)^2}}.
\end{split}
\end{equation}
\begin{equation}
\label{Eq:euclid_similarity_connect}
\begin{split}  
&\hat{r}_{c_1,c_2}=\frac{1}{1+d}=\frac{1}{1+\sqrt{(y_{c_1}-y_{c_2})^{T}(y_{c_1}-y_{c_2})}}\\&=
\frac{1}{1+\sqrt{y_{c_1}^{T}y_{c_1}-y_{c_2}^{T}y_{c_1}-y_{c_1}^{T}y_{c_2}+y_{c_2}^{T}y_{c_2})}}\\&\approx
\frac{1}{1+\sqrt{N(P(c_1=1)+P(c_2=1)-2P(c_1=1,c_2=1))}}.
\end{split}
\end{equation}
\begin{table*}[h]
\setlength{\tabcolsep}{0.8em}
\caption{Comparisons of different similarity metrics (within-dataset protocol).}
\label{tab:sim_within}
\color{black}
\begin{tabular*}{\hsize}{@{}@{\extracolsep{\fill}}lcccc@{}}
\toprule
Methods   & G12EC  & PTB & Ningbo & Chapman\\
\midrule
\multicolumn{5}{c}{\textbf{Ranking loss} (The smaller, the better)}\\
\midrule
Cosine similarity & 0.140$\pm$0.006&0.134$\pm$0.003&0.045$\pm$0.002&0.052$\pm$0.002\\
Pearson coefficient & 0.150$\pm$0.010&0.134$\pm$0.003&0.051$\pm$0.003&0.059$\pm$0.002\\
Euclidean distance  & 0.143$\pm$0.004&0.133$\pm$0.002&0.054$\pm$0.002&0.057$\pm$0.003\\
\midrule
\multicolumn{5}{c}{\textbf{Hamming loss} (The smaller, the better)}\\
\midrule
Cosine similarity & 0.278$\pm$0.008&0.233$\pm$0.009&0.122$\pm$0.001&0.139$\pm$0.002\\
Pearson coefficient & 0.285$\pm$0.010&0.240$\pm$0.008&0.129$\pm$0.002&0.146$\pm$0.007\\
Euclidean distance  & 0.283$\pm$0.004&0.239$\pm$0.006&0.131$\pm$0.003&0.146$\pm$0.007\\
\midrule
\multicolumn{5}{c}{\textbf{Coverage} (The smaller, the better)}\\
\midrule
Cosine similarity & 2.173$\pm$0.027&1.922$\pm$0.015&1.724$\pm$0.010&1.761$\pm$0.021\\
Pearson coefficient & 2.209$\pm$0.034&1.920$\pm$0.012&1.751$\pm$0.014&1.794$\pm$0.012\\
Euclidean distance  & 2.187$\pm$0.013&1.917$\pm$0.011&1.768$\pm$0.008&1.784$\pm$0.023\\
\midrule
\multicolumn{5}{c}{\textbf{MAP} (The greater, the better)}\\
\midrule
Cosine similarity & 0.742$\pm$0.005&0.748$\pm$0.009&0.808$\pm$0.001&0.775$\pm$0.014\\
Pearson coefficient & 0.733$\pm$0.007&0.744$\pm$0.009&0.797$\pm$0.004&0.759$\pm$0.010\\
Euclidean distance  & 0.737$\pm$0.004&0.746$\pm$0.011&0.794$\pm$0.006&0.755$\pm$0.003\\
\midrule
\multicolumn{5}{c}{\textbf{Marco AUC} (The greater, the better)}\\
\midrule
Cosine similarity & 0.854$\pm$0.003&0.880$\pm$0.005&0.925$\pm$0.001&0.912$\pm$0.002\\
Pearson coefficient & 0.851$\pm$0.005&0.880$\pm$0.004&0.915$\pm$0.004&0.907$\pm$0.002\\
Euclidean distance  & 0.855$\pm$0.004&0.882$\pm$0.005&0.915$\pm$0.004&0.907$\pm$0.002\\
\midrule
\multicolumn{5}{c}{\textbf{Marco $G_{beta}$ score} (The greater, the better)}\\
\midrule
Cosine similarity & 0.477$\pm$0.003&0.467$\pm$0.009&0.563$\pm$0.001&0.554$\pm$0.009\\
Pearson coefficient & 0.469$\pm$0.008&0.465$\pm$0.006&0.547$\pm$0.001&0.545$\pm$0.016\\
Euclidean distance  & 0.474$\pm$0.003&0.466$\pm$0.008&0.544$\pm$0.002&0.543$\pm$0.017\\
\bottomrule
\end{tabular*}
\end{table*}

\begin{table*}[h]
\setlength{\tabcolsep}{0.8em}
\begin{center}
\caption{Comparisons of different similarity metrics (cross-dataset protocol)}
\label{tab:sim_cross}
\scalebox{1}{
\color{black}
\begin{tabular*}{\hsize}{@{}@{\extracolsep{\fill}}lcccc@{}}
\toprule
Methods   & G12EC  & PTB & Ningbo & Chapman\\
\midrule
\multicolumn{5}{c}{\textbf{Ranking loss} (The smaller, the better)}\\
\midrule
Cosine similarity & 0.203$\pm$0.004&0.248$\pm$0.005&0.102$\pm$0.006&0.068$\pm$0.002\\
Pearson coefficient & 0.202$\pm$0.003&0.234$\pm$0.007&0.115$\pm$0.013&0.078$\pm$0.004\\
Euclidean distance  & 0.203$\pm$0.007&0.243$\pm$0.005&0.098$\pm$0.003&0.088$\pm$0.003\\
\midrule
\multicolumn{5}{c}{\textbf{Hamming loss} (The smaller, the better)}\\
\midrule
Cosine similarity & 0.331$\pm$0.007&0.310$\pm$0.001&0.253$\pm$0.008&0.219$\pm$0.003\\
Pearson coefficient & 0.335$\pm$0.002&0.320$\pm$0.012&0.281$\pm$0.010&0.249$\pm$0.005\\
Euclidean distance  & 0.328$\pm$0.007&0.331$\pm$0.014&0.276$\pm$0.010&0.262$\pm$0.003\\
\midrule
\multicolumn{5}{c}{\textbf{Coverage} (The smaller, the better)}\\
\midrule
Cosine similarity & 2.415$\pm$0.016&2.379$\pm$0.023&1.971$\pm$0.025&1.803$\pm$0.008\\
Pearson coefficient & 2.415$\pm$0.010&2.324$\pm$0.029&2.027$\pm$0.046&1.852$\pm$0.014\\
Euclidean distance  & 2.422$\pm$0.025&2.364$\pm$0.018&1.963$\pm$0.017&1.898$\pm$0.016\\
\midrule
\multicolumn{5}{c}{\textbf{MAP} (The greater, the better)}\\
\midrule
Cosine similarity & 0.657$\pm$0.009&0.591$\pm$0.012&0.689$\pm$0.002&0.748$\pm$0.004\\
Pearson coefficient & 0.650$\pm$0.002&0.586$\pm$0.006&0.670$\pm$0.003&0.742$\pm$0.005\\
Euclidean distance  & 0.651$\pm$0.007&0.594$\pm$0.009&0.671$\pm$0.004&0.729$\pm$0.001\\
\midrule
\multicolumn{5}{c}{\textbf{Marco AUC} (The greater, the better)}\\
\midrule
Cosine similarity & 0.805$\pm$0.004&0.800$\pm$0.010&0.874$\pm$0.002&0.900$\pm$0.002\\
Pearson coefficient & 0.799$\pm$0.003&0.801$\pm$0.006&0.869$\pm$0.001&0.893$\pm$0.005\\
Euclidean distance  & 0.801$\pm$0.002&0.802$\pm$0.006&0.869$\pm$0.001&0.887$\pm$0.001\\
\midrule
\multicolumn{5}{c}{\textbf{Marco $G_{beta}$ score} (The greater, the better)}\\
\midrule
Cosine similarity & 0.403$\pm$0.002&0.369$\pm$0.001&0.442$\pm$0.003&0.516$\pm$0.006\\
Pearson coefficient & 0.397$\pm$0.002&0.368$\pm$0.003&0.430$\pm$0.005&0.490$\pm$0.006\\
Euclidean distance  & 0.399$\pm$0.004&0.368$\pm$0.005&0.432$\pm$0.005&0.481$\pm$0.005\\
\bottomrule
\end{tabular*}
}
\end{center}
\end{table*}

\begin{table*}[h]
\begin{center}
\caption{Comparisons of different similarity metrics (mix-dataset protocol).}
\label{tab:sim_mix}
\scalebox{1}{
\color{black}
\begin{tabular*}{\hsize}{@{}@{\extracolsep{\fill}}lcccccc@{}}
\toprule
Methods & Ranking loss & Hamming loss & Coverage & MAP & Marco AUC & Marco $G_{beta}$ score \\
\midrule
Cosine similarity & 0.150$\pm$0.001&0.270$\pm$0.001&2.101$\pm$0.009&0.658$\pm$0.006&0.838$\pm$0.003&0.442$\pm$0.002\\
Pearson coefficient & 0.148$\pm$0.007&0.269$\pm$0.007&2.098$\pm$0.037&0.659$\pm$0.003&0.840$\pm$0.002&0.444$\pm$0.003\\
Euclidean distance &0.157$\pm$0.004&0.282$\pm$0.009&2.137$\pm$0.015&0.647$\pm$0.004&0.834$\pm$0.003&0.435$\pm$0.004\\
\bottomrule
\end{tabular*}
}
\end{center}
\end{table*}
\vspace{\fill} 

\ifCLASSOPTIONcaptionsoff
  \newpage
\fi

\end{document}